# Accelerating flux balance calculations in genome-scale metabolic models by localizing the application of loopless constraints

Siu H. J. Chan, Lin Wang, Satyakam Dash, Costas D. Maranas*

Department of Chemical Engineering, The Pennsylvania State University

*Corresponding author (contact: costas@psu.edu)

**Abstract**

**Background**: Genome-scale metabolic network models and constraint-based modeling techniques have become important tools for analyzing cellular metabolism. Thermodynamically infeasible cycles (TICs) causing unbounded metabolic flux ranges are often encountered. TICs satisfy the mass balance and directionality constraints but violate the second law of thermodynamics. Current practices involve implementing additional constraints to ensure not only optimal but also loopless flux distributions. However, the mixed integer linear programming problems required to solve become computationally intractable for genome-scale metabolic models.

**Results**: We aimed to identify the fewest needed constraints sufficient for optimality under the loopless requirement. We found that loopless constraints are required only for the reactions that share elementary flux modes representing TICs with reactions that are part of the objective function. We put forth the concept of localized loopless constraints (LLCs) to enforce this minimal required set of loopless constraints. By combining with a novel procedure for minimal null-space calculation, the computational time for loopless flux variability analysis is reduced by a factor of 10-150 compared to the original loopless constraints and by 4-20 times compared to the currently fastest method Fast-SNP with the percent improvement increasing with model size. Importantly, LLCs offer a scalable strategy for loopless flux calculations for multi-compartment/multi-organism models of very large sizes (e.g. >$10^4$ reactions) not feasible before.

**Introduction**

A genome-scale model (GSM) provides an inventory of reactions for a given organism that allows for the analysis of cellular metabolism and the design of gene modulation strategies for bioproduction. Despite extensive manual curations, thermodynamically infeasible cycles (TIC) often exist in GSMs because of overly permissive reaction inclusion or directionalities that can affect flux range calculations using flux balance analysis (Orth *et al.*, 2010) and/or flux variability analysis (Mahadevan and Schilling, 2003; Gudmundsson and Thiele, 2010). A TIC is an internal cycle in the metabolic network satisfying mass balances and directionality constraints without involving any exchange reactions, i.e., no system input/output is required. For example, a TIC is formed by the following three reactions:

(i) $H_2O$ + Glutamate + $NAD^+$ ↔ $H^+$ + $NH_4^+$ + NADH + 2-Oxoglutarate
(ii) Alanine + 2-Oxoglutarate ↔ Pyruvate + Glutamate
(iii) $H_2O$ + Alanine + $NAD^+$ ↔ $H^+$ + Pyruvate + $NH_4^+$ + NADH

The cycle can carry an arbitrarily large flux when performing FBA despite the fact that any turn around the cycle does not produce or consume any metabolites. Thus, it violates the second law of thermodynamics as the net change of Gibbs free energy is zero. FVA of the TIC participating reactions always predicts unbounded flux ranges. The unbounded fluxes may affect regulatory constraints by making the regulated fluxes essentially unresponsive to any imposed regulation (Dash *et al.*, 2014). In addition, strain design algorithms such as OptForce (Ranganathan *et al.*, 2010; Chowdhury *et al.*, 2014) which rely on precise flux range calculations can also be adversely affected by the presence of TICs.

While TICs can be eliminated by simply *a priori* restricting the directionality of some reactions participating in TICs, this may also rule out biologically realistic phenotypes. In the aforementioned example, thermodynamics dictates that all three reactions are reversible under standard cellular concentrations. Several methods have been proposed to identify and eliminate TICs without over-restricting directionalities. The minsum flux procedure first identifies the maximum biomass yield and then imposes this as a requirement while minimizing the sum of fluxes in the network. A variation of this principle is adopted in parsimonious FBA (Lewis *et al.*, 2010). The minsum flux criterion, however, can become too restrictive ruling out other possibly physiologically meaningful flux distributions and does not necessarily eliminate all TICs. Thermodynamic metabolic flux analysis is a method which takes metabolite concentrations into account to determine directionality, but relies on *a prior* knowledge of the standard Gibbs free energy and physiologically relevant ranges for metabolite concentrations (Henry *et al.*, 2007). Post-processing TIC removal has also been proposed in the recent CycleFreeFlux framework (Desouki *et al.*, 2015). CycleFreeFlux detects and removes TICs in a given flux distribution by solving an additional linear programming (LP) problem and uses an iterative algorithm to calculate FVA ranges for reactions participating in TICs. However, the optimality of the flux distribution with respect to the objective function after TIC removal cannot be guaranteed. Alternatively, loopless FVA (ll-FVA) implementing the loopless constraints directly computes the flux range for a reaction in the absence of any TICs (Schellenberger *et al.*, 2011; Maranas and Zomorrodi, 2016):

$$\min / \max \quad v_j$$

subject to
$$\sum_{j \in \mathbf{J}} S_{ij} v_j = 0, \quad \forall i \in \mathbf{I} \quad \ldots\ldots\ldots(1)$$

$$LB_j \leq v_j \leq UB_j, \quad \forall j \in \mathbf{J} \quad \ldots\ldots\ldots(2)$$

$$-M(1-y_j) \leq v_j \leq My_j, \quad \forall j \in \mathbf{J}^{int} \quad \ldots\ldots\ldots(3)$$

$$-My_j + \varepsilon(1-y_j) \leq G_j \leq -\varepsilon y_j + M(1-y_j), \quad \forall j \in \mathbf{J}^{int} \quad \ldots\ldots\ldots(4)$$

$$\sum_{j \in \mathbf{J}^{int}} N^{int}_{jr} G_j = 0, \quad \forall r \in \{1,\ldots,R\} \quad \ldots\ldots\ldots(5)$$

$$y_j \in \{0,1\}, \quad \forall j \in \mathbf{J}^{int}$$

$$G_j \in \mathbb{R}, \quad \forall j \in \mathbf{J}^{int}$$

$$v_j \in \mathbb{R}, \quad \forall j \in \mathbf{J}$$

where $\mathbf{I}$ is the set of metabolites, $\mathbf{J}$ is the set of reactions, $\mathbf{J}^{int}$ is the set of internal reactions (non-exchange reactions), $\mathbf{S} = [S_{ij}]_{i \in \mathbf{I}, j \in \mathbf{J}}$ is the stoichiometric matrix of the network, $v_j$ is the flux of reaction $j$, $LB_j$ and $UB_j$ are the lower and upper bounds respectively, $\mathbf{N}^{int} = [N_{jr}^{int}]_{j \in \mathbf{J}, r \in \{1,\ldots,R\}}$ is a null-space matrix of rank $R$ of the stoichiometric matrix of internal reactions $\mathbf{S}^{int} = [S_{ij}]_{i \in \mathbf{I}, j \in \mathbf{J}^{int}}$, $G_j$ is a continuous variable associated with internal reaction $j$ analogous to the change in Gibbs free energy of the reaction, $M$ is a large positive constant, $\varepsilon$ is a small positive constant, $y_j$ is a binary variable. ll-FVA formulates a mixed integer linear programming (MILP) problem with the number of binary variables equal to the number of internal reactions of a GSM. It tends to be more time consuming for models of larger sizes, e.g., 10 hours for the *E. coli* iJO1366 model (Orth *et al.*, 2011) as reported in Saa and Nielsen (2016). Saa and Nielsen (2016) presented a novel approach to largely reduce the time for solving FBA/FVA with loopless constraints by invoking a Fast-Sparse Null-space Pursuit (Fast-SNP) algorithm to select a minimal sparse null-space basis $\mathbf{N}^{int}$ that spans a subspace containing all possible internal loops. The reason for the large reduction is that even though the original internal null-space usually has a rank in the order of the number of metabolites in a model, a large part of the null-space does not actually describe any internal loops because the directionality of one or more reactions is violated. Thus, finding the smallest subspace that contains all internal loops satisfying the directionality constraints is sufficient to implement the loopless constraints. Fig. 1 shows an example with three TICs are R1 + R2, R5 + R6 and R5 + R7 (Fig. 1A). The null-space matrix contains a basis vector for R3 + R4 as they can form a cycle considering only the stoichiometry of the two reactions (Fig. 1B). Fast-SNP ruled out this cycle by adding the reaction directionalities (Fig. 1C). In this way, variables $G_j$ and $y_j$ become uncoupled from the rest of the formulation and can be pre-calculated for all reactions $j$ that do not participate in any TIC, for example, $G_{R3}$ and $G_{R4}$, as the coefficients for rows R3, R4 are zero in the null-space matrix. This reduces the number of binary variables from seven to five. One may also apply other approaches to identify all reactions participating in TICs and use the associated null-space for the constraints for loopless solutions. For example, with all the external reactions inactivated we can perform FVA and select all reactions with non-zero fluxes to calculate the null-space of TICs. Using this null-space the number of binary variables being active in ll-FVA (the key determinant of the MILP complexity) can be reduced to the number of reactions participating in any TICs. This concept was recently used to create a loop-free network, instead of performing ll-FVA (Dash *et al.*, 2014).

**Fig. 1. Toy network for illustrating the idea of localized loopless constraints.** (A) A toy network with TICs and the associated stoichiometric matrix. (B) The original constraints for loopless flux calculations imposed on all internal reactions. (C) The previously proposed Fast-SNP (Saa and Nielsen, 2016) to find a minimal null-space. $G_{R3}$, $G_{R4}$, $y_{R3}$, $y_{R4}$ become uncoupled from the rest of the formulation and can be pre-calculated. (D) The proposed LLCs using elementary flux modes that represent TICs. Constraints are imposed only on reactions that are connected to the target reaction R7 by any EFM. The number of binary variables is reduced to two. (E) The proposed LLCs based on the connected components of the null-space. The number of binary variables is three, equal to the size of the connected component containing the target reaction R7. EFM calculations are not required.

In addition to null-space reduction, it is intuitively true that constraints for loopless solutions are redundant while performing FVA analysis for reactions not present in TICs (e.g., R3 in Fig. 1A). Moreover, when finding the flux range for a reaction in a TIC (e.g., R7 of TIC R5 + R7 in Fig. 1A), the constraints for restricting other independent TICs (e.g., R1+R2 in Fig. 1A) can also be removed to simplify the model. As a result, a complete set of loopless constraints defined in ll-FVA is not necessary for analyzing every reaction. Herein, we formalize and generalize these observations by introducing the concept of localized loopless constraints (LLCs) along with a novel algorithm to compute a minimal null-space basis. LLCs are constraints that are only invoked for reactions present in the objective function. We prove that the minimum number of the required binary variables to enforce a loopless requirement when maximizing or minimizing reaction $j$ is equal to the number of reactions that share an elementary flux mode of a TIC with reaction $j$. We prove that as long as the reactions in the objective function and ATP maintenance (ATPM, usually the only reaction with an active lower bound) are not in any TICs, the optimum solution value is unaffected without the constraints for loopless solutions. By using LLCs and a novel null-space algorithm, we are able to further accelerate loopless flux calculations significantly. The null-space calculation time is reduced by 10~1000 times compared to the current available fastest procedure Fast-SNP (Saa and Nielsen, 2016). For the models previously tested using Fast-SNP, LLCs exhibits an improvement of 4 to 10 times in the overall computational time compared to Fast-SNP and 10 to 150 times compared to the use of the original constraints (3) – (4) for loopless solutions. This implies that LLCs is a tractable and scalable strategy for loopless flux calculations in multi-compartment/multi-organism models. We tested community models consisting of multiple *E. coli* and observed 8~20-fold improvement by using LLCs compared to Fast-SNP. The Matlab functions for the COBRA Toolbox (Heirendt *et al.*, 2017) are available in https://github.com/maranasgroup/lll-FVA.

**Methods**
In this section, the concept and applications of LLCs for finding loopless flux distributions given a minimal null-space matrix characterizing all TICs are developed in section 2.1 – 2.4. A novel algorithm for computing a minimal null-space follows in section 2.5. The overall procedure is summarized in section 2.6. Throughout this work, a flux distribution $\mathbf{v} = [v_j]_{j \in \mathbf{J}}$ is defined as a vector satisfying the steady-state condition eq. (1) and the bound constraint eq. (2). $\mathbf{v}^{int} = [v_j]_{j \in \mathbf{J}^{int}}$ denotes the vector containing the fluxes of all internal reactions in $\mathbf{J}^{int}$. The proofs for the propositions presented are given in SI Methods. The network in Fig. 1 is used as a running example to explain the concepts and definitions.

Thermodynamically infeasible cycles
Thermodynamically infeasible cycles (TICs) are defined as follows:

> **Definition 1**
> A TIC is a nonzero flux distribution $\mathbf{v}$ such that $\mathbf{S}^{int}\mathbf{v}^{int} = \mathbf{0}$.

In the toy network, for example, $v_{R1} = v_{R2} = 1$ is a TIC and $v_{ExA} = v_{R1} = v_{R3} = v_{R5} = v_{ExD} = 1$ is not. We now define the relation of a flux distribution containing a sub flux distribution.

**Definition 2**
A flux distribution $\mathbf{v}$ is said to *contain* another flux distribution $\mathbf{v}'$, denoted by $\mathbf{v}' \preceq \mathbf{v}$, if each flux $v'_j$ in $\mathbf{v}'$ is either zero or has the same sign as $v_j$ with a small or equal magnitude, i.e.,
$$0 \leq \text{sgn}(v_j) v'_j \leq \text{sgn}(v_j) v_j, \quad \forall j \in \mathbf{J}$$
where $\text{sgn}(v_j) = 1$ if $v_j \geq 0$ and $= -1$ if $v_j < 0$. Furthermore, $\mathbf{v}$ is said to *properly contain* $\mathbf{v}'$, denoted by $\mathbf{v}' \prec \mathbf{v}$, if in addition to $\mathbf{v}' \preceq \mathbf{v}$ there is at least one reaction carrying nonzero flux in $\mathbf{v}$ but zero flux in $\mathbf{v}'$, i.e.,
$$\exists j \in \mathbf{J} \text{ s.t. } v'_j = 0 \text{ and } v_j \neq 0$$

The relation $\mathbf{v}' \preceq \mathbf{v}$ used in this article has a meaning different from the common usage of $\mathbf{v}'$ being component-wise less than v. Instead, in the context of this work, it implies that $\mathbf{v}'$ and $\mathbf{v}$ have the same dimensions and the magnitude of all fluxes in $\mathbf{v}'$ are between zero and the corresponding value in $\mathbf{v}$. The relation $\mathbf{v}' \prec \mathbf{v}$ implies that in addition to $\mathbf{v}' \preceq \mathbf{v}$ as defined, $\mathbf{v}'$ and $\mathbf{v}$ also satisfy that the set of reactions with zero fluxes for $\mathbf{v}$ is a proper subset of the set of reactions with zero fluxes for $\mathbf{v}'$. We also invoke the concept of elementary flux modes (EFMs), which are flux distributions not properly containing any nonzero flux distribution (Schuster and Hilgetag, 1994).

**Definition 3**
A flux distribution $\mathbf{e}$ is an EFM if $\mathbf{v} \prec \mathbf{e} \Rightarrow \mathbf{v} = \mathbf{0}$.

In other words, a flux distribution $\mathbf{e}$ is an EFM then the only flux distribution it properly contains is $\mathbf{v} = \mathbf{0}$.

Localized loopless constraints
In this subsection, we present the localized loopless constraints (LLCs) for inactivating TICs that involve a specific subset of reactions. Let $\mathbf{J}^{tic} = \{j \mid \exists \text{ TIC } \mathbf{v} \text{ such that } v_j \neq 0\}$ be the set of reactions participating in any TICs (e.g., $\mathbf{J}^{tic} = \{R1, R2, R5, R6, R7\}$ in the toy network of Fig. 1A). Let $\mathbf{K} = \{1, \ldots, K\}$ be the index set for all EFMs $\mathbf{e}_1, \ldots, \mathbf{e}_K$, $\mathbf{K}^{ll} = \{k \in \mathbf{K} \mid \exists j \in \mathbf{J} \setminus \mathbf{J}^{int} \text{ such that } e_{jk} \neq 0\}$ be the set of loopless EFMs. $\mathbf{K}^{tic} = \mathbf{K} \setminus \mathbf{K}^{ll}$ is then the set of all TIC EFMs (see Fig. 1D). Assume that only specific TICs involving a set of target reactions in $\mathbf{T} \subset \mathbf{J}^{tic}$ are required to be inactivated in the flux distribution, i.e., TICs not containing any reactions in $\mathbf{T}$ can still be part of the flux distribution. Denote the set of TIC EFMs involving any reactions in **T** by $\mathbf{K}_\mathbf{T}^{tic} = \{k \in \mathbf{K}^{tic} \mid \exists j \in \mathbf{T} \text{ such that } e_{jk} \neq 0\}$. By the 'no-cancellation' rule (Schuster *et al.*, 2002) used extensively before (Schwartz and Kanehisa, 2005, 2006; Zhao and Kurata, 2009; Carlson, 2009; Chan and Ji, 2011; Ip *et al.*, 2011; Chan *et al.*, 2014), any flux distribution **v** can be decomposed into EFMs as follows:

$$\begin{aligned} \mathbf{v} &= \mathbf{v}^{ll} + \mathbf{v}^{tic} \\ &= \mathbf{v}^{ll} + \mathbf{v}^{tic,\,nontarget} + \mathbf{v}^{tic,\,target} \\ &= \sum_{k \in \mathbf{K}^{ll}} \alpha_k \mathbf{e}_k + \sum_{k \in \mathbf{K}^{tic} \setminus \mathbf{K}_\mathbf{T}^{tic}} \alpha_k \mathbf{e}_k + \sum_{k \in \mathbf{K}_\mathbf{T}^{tic}} \alpha_k \mathbf{e}_k, \quad \alpha_k \geq 0, \, \forall k \in \mathbf{K} \end{aligned} \quad (6)$$

where $\mathbf{v}^{ll}$ is the loopless part of the flux distribution **v** and $\mathbf{v}^{tic}$ consists of TICs only. $\mathbf{v}^{tic}$ can be decomposed into TICs not involving any target reactions in **T** ($\mathbf{v}^{tic,\,nontarget}$) and TICs through reactions in **T** ($\mathbf{v}^{tic,\,target}$). From eq. (6), blocking the target set of EFMs $\mathbf{K}_\mathbf{T}^{tic}$ is sufficient to ensure $\mathbf{v}^{tic,\,target} = \mathbf{0}$ and thus eliminate TICs through reactions in **T** from **v**. Let $\mathbf{C}_\mathbf{T}^{EFM} = \{j \in \mathbf{J}^{tic} \mid \exists k \in \mathbf{K}_\mathbf{T}^{tic} \text{ such that } e_{jk} \neq 0\}$ be the set of reactions connected to **T** by any target TIC EFMs. It contains all reactions that can have nonzero fluxes in $\mathbf{v}^{tic,\,target}$. In the toy network, R7 is used as an example target reaction, i.e., **T** = {R7}. From the EFM matrix, only the fifth EFM is a TIC EFM that involves R7, therefore, $\mathbf{K}_\mathbf{T}^{tic}$ ={5} and $\mathbf{C}_\mathbf{T}^{EFM}$ ={R5, R7} (see Figure 1E). Proposition 1 states that the LLCs imposed on $\mathbf{C}_\mathbf{T}^{EFM}$, which are eqs. (7) – (8), eliminate TICs involving **T**.

**Proposition 1**
A flux distribution **v** does not contain any TICs involving reactions in **T** if eq. (3) and eqs. (4) – (5) restricted to $j \in \mathbf{C}_\mathbf{T}^{EFM}$ are satisfied:

$$-M(1-y_j) \leq v_j \leq My_j, \qquad \forall j \in \mathbf{C}_\mathbf{T}^{EFM} \quad (7)$$

$$-My_j + \varepsilon(1-y_j) \leq G_j \leq -\varepsilon y_j + M(1-y_j), \quad \forall j \in \mathbf{C}_\mathbf{T}^{EFM} \quad (8)$$

$$\sum_{j \in \mathbf{J}^{int}} N_{jr}^{int} G_j = 0, \qquad \forall r \in \{1,\dots,R\} \quad (9)$$

$$g_j \in \mathbb{R}, \qquad \forall j \in \mathbf{J}^{int}$$

$$y_j \in \{0,1\}, \qquad \forall j \in \mathbf{C}_\mathbf{T}^{EFM}$$

See SI Methods for the proof. The idea is to derive the equations by constraining $\mathbf{v}^{tic,\,target}$ in eq. (6) to zero. The LLCs and the proof in SI Methods refine the original constraints and the previously presented proof (Noor *et al.*, 2012). In the toy network, by Proposition 1, eqs. (7) – (8) constraining only {R5, R7} are sufficient to prevent any TIC involving R7. However, if **T** = {R5}, since there is one TIC EFM connecting R5 and R6 and one connecting R5 and R7, $\mathbf{C}_\mathbf{T}^{EFM}$ ={R5, R6, R7} is required to be constrained in eqs. (7) – (8) to prevent TICs involving R5.

Determination of the target reaction set for applying LLCs

Proposition 1 implies that the number of binary variables can be reduced if only the specific TICs involving the target reactions in **T** (e.g., R5 and R7 in the toy network) are required to be inactivated. Indeed in many applications, the optimal objective function value of a MILP with the original constraints for loopless solutions is equal to that with suitably selected LLCs. The optimal loopless solution can also be derived from the corresponding partially loopless solution. Consider the following general LP problem for finding a flux distribution **v**:

$$\min_{v_j, j \in \mathbf{J}} \quad \sum_{j \in \mathbf{J}} c_j v_j$$
$$\text{subject to} \quad \sum_{j \in \mathbf{J}} S_{ij} v_j = 0, \quad \forall i \in \mathbf{I}$$
$$\sum_{j \in \mathbf{J}} a_{pj} v_j \leq b_p \quad \forall p \in \{1, \ldots, P\} \quad (10)$$

where $P$ is the number of additional constraints. From eq. (6), by applying LLCs, we already have $\mathbf{v}^{\text{tic,target}} = \mathbf{0}$. Obviousely, both $\mathbf{v}^{\text{ll}}$ and $\mathbf{v}^{\text{tic,nontarget}}$ satisfy the mass balance equation. Therefore, as long as $\mathbf{v}^{\text{tic,nontarget}}$ does not contribute to the optimal objective function value (i.e., $\sum_{j \in \mathbf{J}^{\text{tic}}} c_j v_j^{\text{tic, nontarget}} = 0$) and $\mathbf{v}^{\text{ll}}$ alone satisfies the additional $P$ constraints (i.e., $\sum_{j \in \mathbf{J}} a_{pj} v_j^{\text{ll}} \leq b_p$ for all $p$), then $\mathbf{v}^{\text{ll}}$ is an optimal feasible solution to problem (10). Proposition 2 establishes a useful sufficient condition for using LLCs based on this idea.

**Proposition 2**

Denote the sets of reactions whose forward and reverse directions participate in TICs respectively by $\mathbf{J}_{\text{fwd}}^{\text{tic}} = \{j \in \mathbf{J}^{\text{tic}} | e_{jk} \geq 0 \ \forall k \in \mathbf{K}^{\text{tic}}\}$ and $\mathbf{J}_{\text{rev}}^{\text{tic}} = \{j \in \mathbf{J}^{\text{tic}} | e_{jk} \leq 0 \ \forall k \in \mathbf{K}^{\text{tic}}\}$. For the LP problem (10), assume that the target reaction set $\mathbf{T} \subset \mathbf{J}^{\text{tic}}$ contains all reactions in TICs satisfying one of the following three conditions:

(I) $j \in \mathbf{J}_{\text{fwd}}^{\text{tic}}$ and $c_j < 0$

(II) $j \in \mathbf{J}_{\text{rev}}^{\text{tic}}$ and $c_j > 0$

(III) Conditions (A) and (B) are satisfied:

(A) $(j \in \mathbf{J}_{\text{fwd}}^{\text{tic}}$ and $\exists p$ s.t. $a_{pj} < 0)$ or $(j \in \mathbf{J}_{\text{rev}}^{\text{tic}}$ and $\exists p$ s.t. $a_{pj} > 0)$

(B) $\exists p$ s.t. $a_{pj} \neq 0$ and $(a_{pj'} \neq 0$ for some $j' \neq j$ or $b_p < 0)$

Then the optimal objective function value of the LP problem (10) constrained with the localized loopless constraints eqs. (7) – (9) is equal to the optimal objective function value of the LP problem (10) constrained with the original loopless constraints eqs. (3) – (5).

Conditions (I) and (II) state that a reaction in $\mathbf{J}_{\text{fwd}}^{\text{tic}}$ or in $\mathbf{J}_{\text{rev}}^{\text{tic}}$ needs to be put in the target set **T** only when the reaction flux is being maximized or minimized, respectively. In the toy network, $\mathbf{J}_{\text{fwd}}^{\text{tic}} = \{R1, R2, R5, R6, R7\}$ and $\mathbf{J}_{\text{rev}}^{\text{tic}}$ is empty. Though R1 and R5 are reversible, the reverse direction does not participate in any TIC. Condition (III.A) excludes a reaction that is in TICs from **T** even when it is constrained in the problem as long as it participates in TICs only in the forward (or reverse) direction and meanwhile the corresponding constraint coefficients for that reaction are all non-negative (or non-positive). Condition (III.B) excludes most of the bound constraints on fluxes in eq. (2) with $LB_j \leq 0$ and $UB_j \geq 0$. For example, $v_{R2} \leq 1$ and $v_{R2} \geq 0$ do not satisfy conditions (III.A) and (III.B) respectively. These constraints do not necessitate putting R2 in **T**. For $v_{R2} \geq 1$ and $-v_{R2} + v_{R4} \leq 0$, both of them satisfy conditions (III.A) and (III.B). If any one of the two constraints is in the problem, R2 must be included in **T**. The proof is provided in SI Methods. An optimal loopless flux distribution $\mathbf{v}^{ll}$ can be obtained by removing TICs from the localized loopless flux distribution as a post-processing step as proposed in cycleFreeFlux (Desouki et al., 2015). In the toy network, without other constraints, Proposition 3 states that applying LLCs with **T** = {R7} is sufficient to find the maximum of $v_{R7}$. Solving with LLCs may result at the flux distribution:

$$\begin{matrix} R1 & R2 & R3 & R4 & R5 & R6 & R7 & ExA & ExD \\ [1000 & 1000 & 0 & 0 & 1000 & 1000 & 0 & 0 & 0]^T \end{matrix}$$

Despite the fact that it contains TICs, LLCs on R5, R7 prevent any TICs through R7. The completely loopless flux distribution $\mathbf{v} = \mathbf{0}$ can be obtained by post processing such as cycleFreeFlux. In this way, the number of binary variables in solving MILP is reduced from 7 using the original null-space (see Fig. 1B) to 5 using the Fast-SNP null-space (see Fig. 1C), and further down to 2 using the localized loopless constraints (see Fig. 1D). We define the reduction fraction $f$ as the fraction of reduction in binary variables when using LLCs compared to using eqs. (3) – (5) with a minimal null-space, i.e.,

$$f = |\mathbf{C}_{\mathbf{T}}^{\text{EFM}}| / |\mathbf{J}^{\text{tic}}|$$

When applying LLCs to R7, $f = 0.4$. Taking **T** = {R5} as another example, although R5 is in TICs and is reversible, only the forward direction of R5 is in TICs. When minimizing $v_{R5}$ ($c_{R5} = 1$ and $R5 \notin \mathbf{J}_{\text{rev}}^{\text{tic}}$ do not satisfy any of the conditions in Proposition 2), solving the LP problem gives the same minimum for $v_{R5}$ as solving with the original constraints for loopless solutions. In this case, $f = 0$. A common constraint in metabolic models that satisfies condition (III) of Proposition 3 is the positive lower bound for the ATP maintenance (ATPM) reaction. Nonetheless, ATPM appearing in a TIC implies that it is coupled to an energy generating cycle, which should be resolved by manual curation (Fritzemeier *et al.*, 2017). Therefore, Proposition 2 guarantees that in a well-curated model free of ATP-generating cycles, the standard FBA and FVA can be performed without any constraints for loopless solution when the objective function does not concern any reactions in TICs (i.e., **T** is empty).

Finding a superset for reactions connected to the target set

One challenge in implementing LLCs is to determine $\mathbf{C}_\mathbf{T}^{EFM}$, the set of reactions connected to **T** by any TIC EFMs. This entails the use of the entire set of EFMs which in some cases could be computationally intractable due to combinatorial explosion (Klamt and Stelling, 2002). However, the minimal null-space matrix offers a computationally efficient way to find a superset of $\mathbf{C}_\mathbf{T}^{EFM}$ that is smaller than $\mathbf{J}^{tic}$ when computing EFMs is not preferred. Using the minimal null-space matrix, we can define that two reactions are connected if (i) they both have nonzero values in a column in the null-space matrix or (ii) if there is a reaction connected to both of them (and therefore the connectivity is transitive). Fig. 1E visualizes the relation in the toy network. Pairs R5 + R6 and R5 + R7 satisfy condition (i). Therefore, R5 is connected to both R6 and R7. By condition (ii), R6 and R7 are also connected via R5. Under this relation of connection, the reactions in TICs $\mathbf{J}^{tic}$ can be partitioned into $L$ connected components $\mathbf{J}_l^{tic} \subset \mathbf{J}^{tic}$ for $l = 1,\ldots,L$ (e.g., $\mathbf{J}_1^{tic} =$ {R1, R2} and $\mathbf{J}_2^{tic} =$ {R5, R6, R7} in Fig. 1E). For any union of connected components $\mathbf{C}_\mathbf{T}^{NS} = \bigcup_{q=1}^{Q} \mathbf{J}_{l_q}^{tic}$ where $1 \leq l_1,\ldots,l_Q \leq L$, we prove that if $\mathbf{C}_\mathbf{T}^{NS}$ contains **T**, then it also contains $\mathbf{C}_\mathbf{T}^{EFM}$ (see SI Methods for proof), i.e.,

$$\mathbf{T} \subset \mathbf{C}_\mathbf{T}^{NS} \Rightarrow \mathbf{C}_\mathbf{T}^{EFM} \subset \mathbf{C}_\mathbf{T}^{NS}$$

Imposing LLCs on $\mathbf{C}_\mathbf{T}^{NS}$ is thus sufficient to ensure the absence of TICs through reactions in **T**. In the toy network, we can apply LLCs on the connected component $\mathbf{C}_\mathbf{T}^{NS} = \mathbf{J}_2^{tic} =$ {R5, R6, R7} to ensure no TICs through R7. In this case, we still have a reduction fraction $f = 0.6$.

Algorithm for minimal null-space

In addition to introducing LLCs, we propose here a novel algorithm for computing a minimal null-space. The current best method Fast-SNP constructs a minimal null-space by iteratively solving LP problems to find a new feasible basis vector not lying in the null-space under construction until no new basis vector is found. Instead, we propose to solve a single MILP problem that find a maximal TIC such that each reaction in $\mathbf{J}^{tic}$ carries nonzero flux. A minimal null-space basis is then calculated from the submatrix $\mathbf{S}^{tic} = [S_{ij}]_{i \in \mathbf{I}, j \in \mathbf{J}^{tic}}$. The procedure recovers the null-space in a significantly shorter time for large models compared to Fast-SNP. By introducing constraints similar to eqs. (3) − (4) to model the flux direction, the MILP problem is formulated as follows:

$$\begin{aligned}
\min \quad & \sum_{j \in \mathbf{J}^{\text{int}}} \left(z_j^+ + z_j^-\right) \\
\text{s.t.} \quad & \sum_{j \in \mathbf{J}^{\text{int}}} S_{ij} v_j = 0, & \forall i \in \mathbf{I} \\
& -M\delta_j^L \leq v_j \leq M\delta_j^U, & \forall j \in \mathbf{J}^{\text{int}} \\
& \varepsilon - M(z_j^+ + 1 - \delta_j^U) \leq v_j \leq -\varepsilon + M(z_j^- + 1 - \delta_j^L), & \forall j \in \mathbf{J}^{\text{int}} \\
& z_j^+, z_j^- \geq 0 & \forall j \in \mathbf{J}^{\text{int}} \\
& v_j \in \mathbb{R} & \forall j \in \mathbf{J}^{\text{int}} \\
& z_j^+, z_j^- \in \{0,1\} & \forall j \in \mathbf{J}^{\text{int}} \text{ if } \delta_j^L = \delta_j^U = 1 \quad (11)
\end{aligned}$$

where $\delta_j^L = 1$ if $LB_j < 0$ and zero otherwise whereas $\delta_j^U = 1$ if $UB_j > 0$ and zero otherwise. Thus, for a reversible reaction $j$, $\delta_j^L = \delta_j^U = 1$. The proof that all reactions in TICs have nonzero fluxes in the solution of problem (11) is provided in SI Methods. Problem (11) forces each reaction in TICs to have a nonzero flux by minimizing $z_j^+ + z_j^-$. $z_j^+ = 0$ implies that reaction $j$ can have positive flux in TICs and $z_j^- = 0$ implies that reaction $j$ can have negative flux. For each irreversible reaction $j$, one of the $z_j^+, z_j^-$ can be predetermined ($z_j^+ = 0$ if $\delta_j^U = 0$ and $z_j^- = 0$ if $\delta_j^L = 0$) and only the other needs to be determined. Modeling $z_j^+, z_j^-$ as continuous variables for an irreversible reaction $j$ suffices to force $v_j \neq 0$ if it is feasible. For a reversible reaction $j$, $z_j^+, z_j^-$ are required to be binary. A special property of problem (11) is that during branch and bound for solving MILP, branching down (take $z_j^+$ or $z_j^- = 0$) whenever possible until integer feasibility can always lead to an optimal solution (see SI Methods for more detailed analysis) implying that at most $2n_{rev}$ relaxed LPs need to be solved for $n_{rev}$ reversible reactions. Solving problem (11) is similar to an iterative LP procedure but it takes advantage of the built-in structure of any modern MILP solver. In practice, problem (11) is pre-solved as LPs to determine $\mathbf{J}_{\text{fwd}}^{\text{tic}}$ and $\mathbf{J}_{\text{rev}}^{\text{tic}}$ simultaneously (see SI Methods).

Overall procedure

Combining all the methods presented, we propose the following procedure for performing loopless flux balance calculations:

1. Solve problem (11) to identify $\mathbf{J}^{\text{tic}}$ and compute a minimal null-space matrix $\mathbf{N}^{\text{int}}$ using the submatrix $\mathbf{S}^{\text{tic}} = [S_{ij}]_{i \in \mathbf{I}, j \in \mathbf{J}^{\text{tic}}}$.

2. Identify the set of reactions whose forward directions are in TICs $\mathbf{J}_{\text{fwd}}^{\text{tic}}$ and whose reverse directions are in $\mathbf{J}_{\text{rev}}^{\text{tic}}$ by FVA on reactions in $\mathbf{J}^{\text{tic}}$ with all exchange reactions shut down.

3. Find all connected components $\mathbf{J}_l^{\text{tic}} \subset \mathbf{J}^{\text{tic}}$ from the null-space.

4. For each loopless flux calculation, determine the target set $\mathbf{T}$ for applying LLCs using the conditions in Proposition 2.

5. Find the minimum $\mathbf{C}_{\mathbf{T}}^{\mathrm{NS}} = \bigcup_{q=1}^{Q} \mathbf{J}_{l_q}^{\mathrm{tic}}$ such that $\mathbf{T} \subset \mathbf{C}_{\mathbf{T}}^{\mathrm{NS}}$. For each $q = 1, \ldots, Q$, calculate the complete set of EFMs for connected component $\mathbf{J}_{l_q}^{\mathrm{tic}}$ using the corresponding columns from the stoichiometric matrix $\mathbf{S}$. Determine $\mathbf{C}_{\mathbf{T}}^{\mathrm{EFM}}$ from the EFMs.

6. Solve problem (10) as an MILP problem by imposing constraints (7) – (9) on $\mathbf{C}_{\mathbf{T}}^{\mathrm{EFM}}$.

See SI Methods for more details on Step 1 – 3. If the calculation of EFMs in Step 5 is not preferable, one can skip the EFM calculation and solve the MILP problem with LLCs imposed on $\mathbf{C}_{\mathbf{T}}^{\mathrm{NS}}$ instead of $\mathbf{C}_{\mathbf{T}}^{\mathrm{EFM}}$. We have implemented the procedure in MATLAB using the COBRA Toolbox (Heirendt *et al.*, 2017) and the optimization solver Gurobi (http://www.gurobi.com). *EFMtool* was used for EFM computations (Terzer and Stelling, 2008).

**Results**

Single-organism models

Using the seven models tested in Saa and Nielsen (2016) excluding the toy model therein, we compared the performance between (1) ll-FVA using the original loopless constraint (referred to as ll-FVA), (2) ll-FVA with Fast-SNP preprocessing (referred to as Fast-SNP), (3) FVA with null-space-based LLCs, i.e., imposing LLCs on $\mathbf{c}_{\mathbf{T}}^{\mathrm{NS}}$ in Step 6 in section 2.6 without EFM calculations (referred to as NS-LLC), and (4) FVA with EFM-based LLCs (referred to as EFM-LLC). In the models tested, NS-LLC and EFM-LLC show 10~150x reduction in computational needs (in CPU time) compared to ll-FVA and 4~10x reduction compared to Fast-SNP (Table 1). Figure 2A shows the breakdown of the computational time into the time allotted for null-space calculation (Step 1), LLC preprocessing (Step 2 - 4), EFM calculation (Step 5, applied to EFM-LLC only) and solving MILP problems (Step 6). For null-space calculation, in all models except the *E. coli* core model, the proposed procedure is 4~15x faster than Fast-SNP. The time for NS-LLC preprocessing is in general small and accounts for only <0.5% of the total CPU time. In contrast, the time for EFM calculation is more model-dependent and accounts for 12% to 64% of the total run time respectively, for the *E. coli* core model and yeast6 models. In contrast, EFM-LLC required an MILP solution time 2.5 times faster than that of NS-LLC for the yeast6 model. To understand how NS-LLC and EFM-LLC greatly shorten the MILP time compared to Fast-SNP, Figure 2B shows the breakdown of the MILP time into the time taken to solve MILP problems with various levels of reduction fraction $f$, defined by the number of 0-1 variables in MILP problems divided by the number of reactions in TICs. For all models except yeast6, a significant amount of the solution time was spent on solving problems without any integer variables (reduction fraction $f = 0$), i.e. LP problems only. For yeast6, NS-LLC spent a significant amount of time on solving MILP problems with $0.5 < f < 1$ while most MILP problems for EFM-LLC have $f < 0.5$. The results demonstrate the effectiveness of NS-LLC and EFM-LLC in reducing MILP complexity.

**Fig. 2. Performance of the four methods for loopless FVA under comparision on single-organism models.** (A) CPU time allotted for minimal null-space calculation (Step 1), LLC preprocessing (Step 2 – 4), EFM computation (Step 5), and MILP solution time (Step 6). (B) Breakdown of the MILP solution time into time for solving problems with various degree of binary variable reduction in terms of the reduction fraction $f$.

Multi-organism models

The computational performance of Fast-SNP, NS-LLC and EFM-LLC was further compared using models for microbial communities. The nine-species model previously used for modeling the gut microbiota (Chan *et al.*, 2017b) and community models consisting of multiple copies of the *E. coli* iJO1366 (Orth *et al.*, 2011) were tested. In the community models, inter-organism TICs are eliminated without the need of over-restricting the directionality of intracellular reactions by imposing suitable penalties on transport reactions, e.g., consuming instead of gaining proton gradient by exporting certain metabolites (Chan *et al.*, 2017a). Otherwise, when the number of organisms increases, extremely large inter-organism TICs (involving hundreds to thousands of reactions) can appear and cause the MILP problems to become intractable. Restricting TICs to appear only within individual organisms and applying LLCs can become a scalable procedure for loopless flux calculations for multi-organism models because the number of binary variables in each MILP problem solved is independent of the number of organisms. Table 2 shows the computational results of loopless FVA for all reactions in TICs. They are the reactions for which non-trivial MILP problems must be solved. Overall, NS-LLC and EFM-LLC show improvements in computation speed of 8~16 times and 8~19 times, respectively, compared to Fast-SNP. The null-space preprocessing time using the proposed procedure compared to Fast-SNP exhibits a ~$10^2$- to ~$10^3$-fold decrease (Fig. 3A). The total MILP solution time using NS-LLC and EFM-LLC is reduced by 8~15 and 8~18 times compared to Fast-SNP, respectively (Fig. 3B). This is caused by the larger extent of binary variable reduction (smaller reduction fraction $f$) as the model size grows (Fig. 3B), confirming the particular advantage of LLCs for multi-organism and multi-compartment models. Fig. 4A and Fig. 4B compare the solution time taken by NS-LLC and EFM-LLC compared to Fast-SNP for the same MILP problems solved, respectively. They clearly show that as the model size increases, the MILP complexity of Fast-SNP grows more rapidly than NS-LLC and EFM-LLC. The large standard deviations signify the fact that the actual reduction in computational time is very problem-specific.

**Fig. 3. Performance of Fast-SNP, NS-LLC and EFM-LLC for loopless FVA on community models.** (A) Breakdown of total CPU time into the time for minimal null-space calculation (Step 1), localized loopless constraint preprocessing (Step 2 – 4), EFM computation (Step 5A), and MILP solution time (Step 6). (B) Breakdown of the MILP solution time into time for solving problems with various degree of binary variable reduction in terms of the reduction fraction $f$. Ec($n$) is the community model of $n$ copies of the *E. coli* iJO1366 model.

**Fig. 4. Comparison of the solution time for individual MILP problems.** (A) The ratio of the MILP solutiontime by Fast-SNP ($t_{Fast\text{-}SNP}^{MILP}$) to that by NS-LLC ($t_{NS-LLC}^{MILP}$). (B) The ratio of the MILP solution time by Fast-SNP to that by EFM-LLC ($t_{EFM\text{-}LLC}^{MILP}$). Each plotted value is the mean ratio of all MILP problems solved for the model and the error bar represents one standard deviation. Ec($n$) is the community model of $n$ copies of the *E. coli i*JO1366 model.

## Conclusions

In this paper, we propose the use of LLCs to further reduce the computational cost for loopless flux calculations. Notably we proved that for many models simply computing the LP problem without loopless constraints is sufficient to guarantee optimality. An optimal and loopless flux distribution can be derived from the LP solution by a simple post-processing TIC removal. LLCs offer a scalable way for loopless flux calculations for large multi-organism/multi-compartment genome-scale metabolic models. Since in most tested models except the yeast6 model, EFM-LLC slightly outperformed NS-LLC, we recommend using EFM-LLC in the initial attempt. If the EFM calculation is intractable, NS-LLC can be used. While NS-LLC guarantees an efficient preprocessing (finding connected components from the null-space matrix), it does not ensure the largest degree of binary variable reduction as EFM-LLC does. However, the cost of EFM computation in EFM-LLC preprocessing can be unpredictably high due to combinatorial explosion (Klamt and Stelling, 2002). The information required from the computed TIC EFMs is only whether two reactions are linked by any TIC EFMs. Computing the entire set of TIC EFMs is in most cases unnecessary. If an efficient algorithm for determining whether a TIC EFM exists between any two reactions in TICs can be devised, then the efficiency of loopless flux calculations can be further improved.

## Funding

This work has been supported by the National Science Foundation (NSF) grant no. NSF/MCB 1546840 and the U.S. Department of Energy (DOE, http://www.energy.gov/) grant no. (DOE DE-SC0012377).

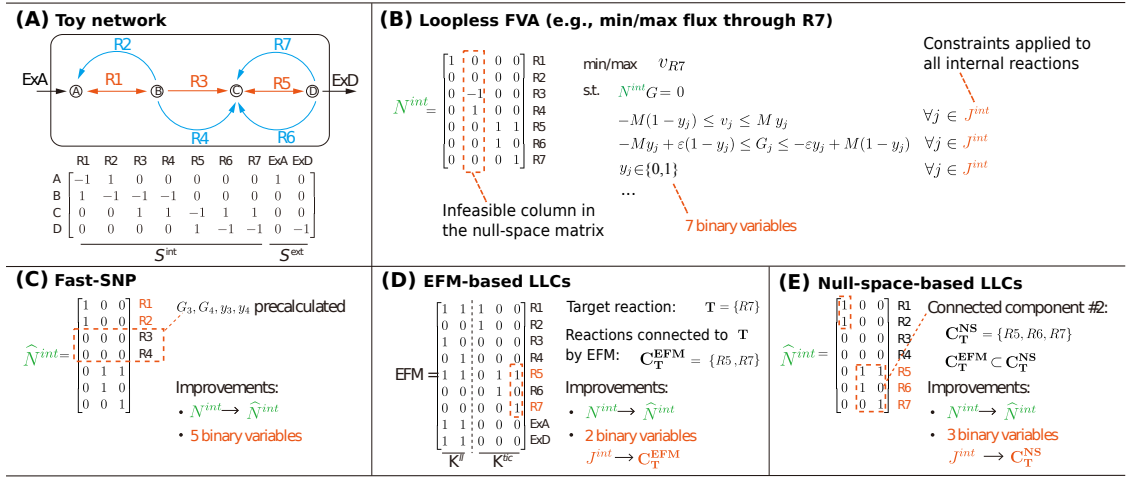

**Fig. 1. Toy network for illustrating the idea of localized loopless constraints.** (A) A toy network with TICs and the associated stoichiometric matrix. (B) The original constraints for loopless flux calculations imposed on all internal reactions. (C) The previously proposed Fast-SNP (Saa and Nielsen, 2016) to find a minimal null-space. $G_{R3}$, $G_{R4}$, $y_{R3}$, $y_{R4}$ become uncoupled from the rest of the formulation and can be pre-calculated. (D) The proposed LLCs using elementary flux modes that represent TICs. Constraints are imposed only on reactions that are connected to the target reaction R7 by any EFM. The number of binary variables is reduced to two. (E) The proposed LLCs based on the connected components of the null-space. The number of binary variables is three, equal to the size of the connected component containing the target reaction R7. EFM calculations are not required.

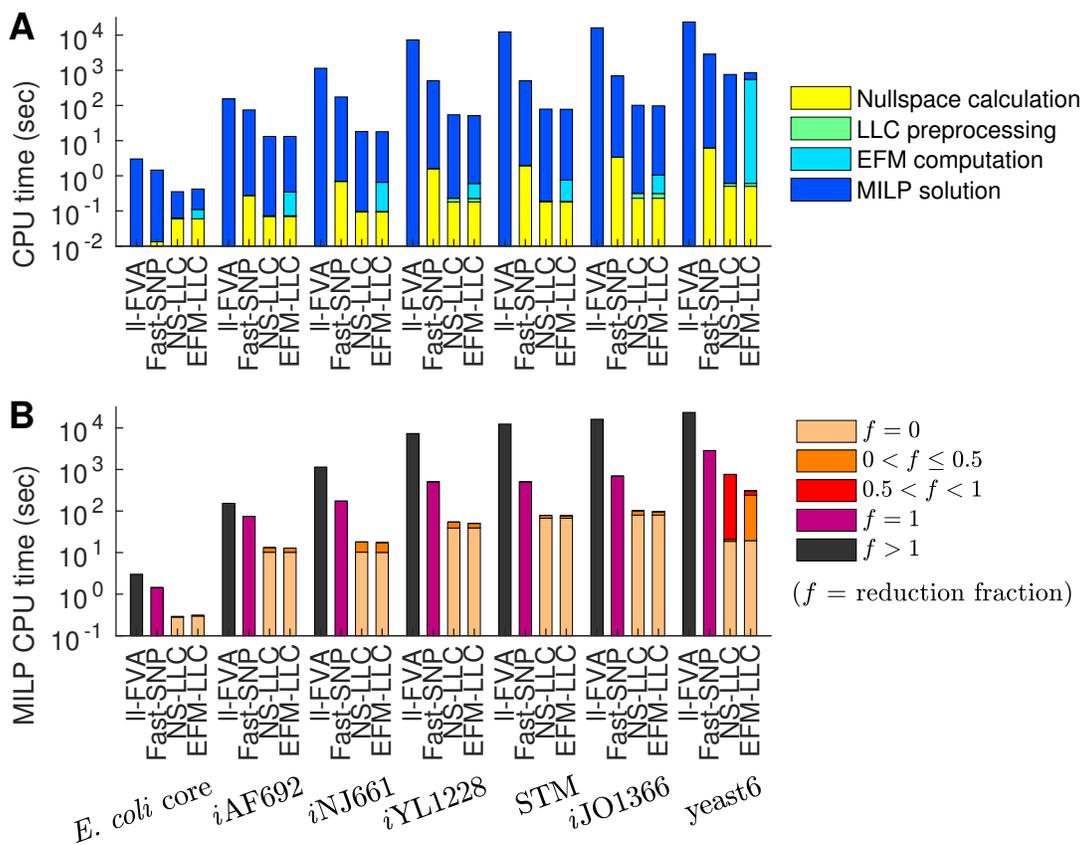

**Fig. 2. Performance of the four methods for loopless FVA under comparision on single-organism models.** (A) CPU time allotted for minimal null-space calculation (Step 1), LLC preprocessing (Step 2 – 4), EFM computation (Step 5), and MILP solution time (Step 6). (B) Breakdown of the MILP solution time into time for solving problems with various degree of binary variable reduction in terms of the reduction fraction $f$.

**Fig. 3. Performance of Fast-SNP, NS-LLC and EFM-LLC for loopless FVA on community models.** (A) Breakdown of total CPU time into the time for minimal null-space calculation (Step 1), localized loopless constraint preprocessing (Step 2 – 4), EFM computation (Step 5A), and MILP solution time (Step 6). (B) Breakdown of the MILP solution time into time for solving problems with various degree of binary variable reduction in terms of the reduction fraction $f$. Ec($n$) is the community model of $n$ copies of the *E. coli* iJO1366 model.

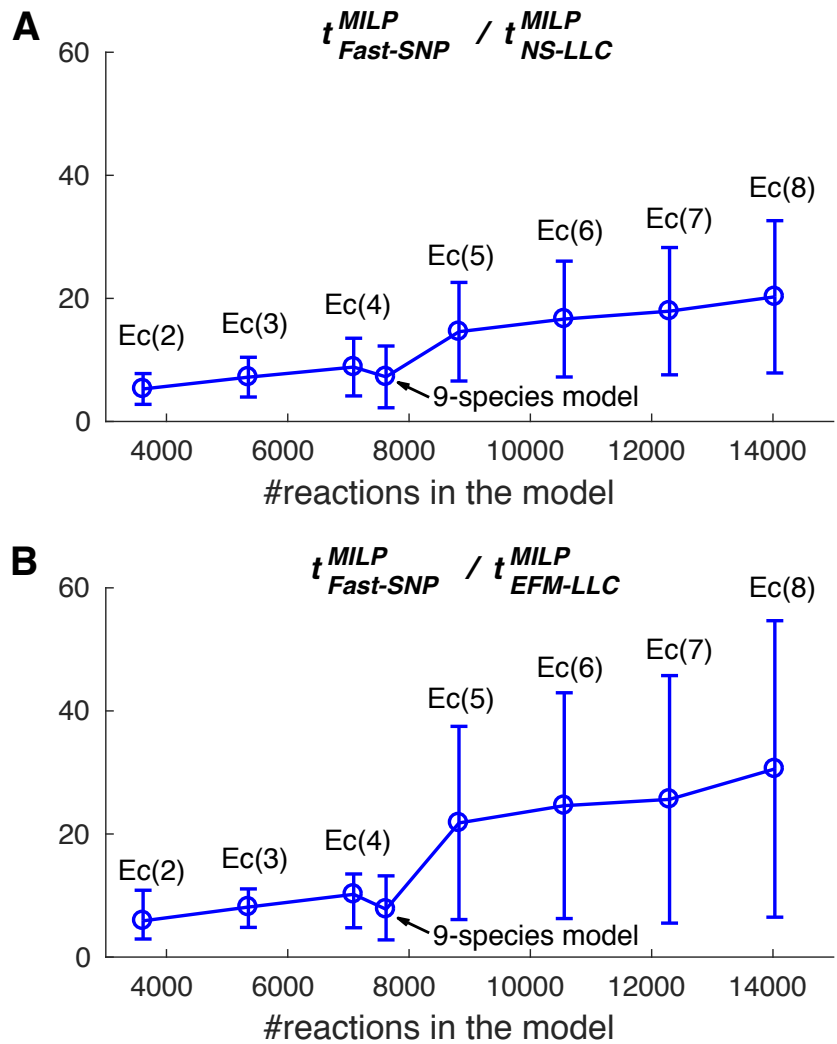

**Fig. 4. Comparison of the solution time for individual MILP problems.** (A) The ratio of the MILP solutiontime by Fast-SNP ($t_{Fast\text{-}SNP}^{MILP}$) to that by NS-LLC ($t_{NS-LLC}^{MILP}$). (B) The ratio of the MILP solution time by Fast-SNP to that by EFM-LLC ($t_{EFM\text{-}LLC}^{MILP}$). Each plotted value is the mean ratio of all MILP problems solved for the model and the error bar represents one standard deviation. Ec($n$) is the community model of $n$ copies of the *E. coli* iJO1366 model.

**Table 1.** Computational time for the four methods for loopless FVA under comparison tested for single-organism models.

| Model | #metabolites* | #reactions* | $|\mathbf{J}^{tic}|$ | Total CPU time (sec) | | | | Max. diff.[&] |
|---|---|---|---|---|---|---|---|---|
| | | | | ll-FVA | Fast-SNP | NS-LLC | EFM-LLC | |
| *E. coli* core | 68 | 87 | 2 | 3 ± 0.09 | 1.46 ± 0.01 | 0.35 ± 0.06 | 0.42 ± 0.05 | $1.9 \times 10^{-10}$ |
| *i*AF692 | 417 | 484 | 30 | 153 ± 12 | 75 ± 5 | 13 ± 0.2 | 13 ± 0.3 | $2.5 \times 10^{-7}$ |
| *i*NJ661 | 579 | 740 | 53 | 1143 ± 51 | 174 ± 12 | 18 ± 1 | 18 ± 1 | $1.0 \times 10^{-7}$ |
| *i*YL1228 | 830 | 1223 | 59 | 7330 ± 549 | 504 ± 24 | 54 ± 2 | 51 ± 1 | $3.3 \times 10^{-7}$ |
| STM | 1086 | 1597 | 52 | 12376 ± 1930 | 503 ± 11 | 79 ± 2 | 78 ± 1 | $4.1 \times 10^{-8}$ |
| *i*JO1366 | 1136 | 1679 | 76 | 16173 ± 2387 | 699 ± 29 | 101 ± 2 | 98 ± 1 | $4.6 \times 10^{-8}$ |
| yeast6 | 756 | 1018 | 293 | 23614 ± 1865 | 2847 ± 105 | 760 ± 68 | 859 ± 118 | $8.0 \times 10^{-8}$ |

$|\mathbf{J}^{tic}|$ is the number of reactions in TICs.
*Numbers of metabolites and reactions after removal of blocked reactions.
[^]Mean ± s.d. of three replications.
[&]The maximum absolute difference of flux values between all tested methods and ll-FVA for each model.

**Table 2**. Computational time for the tested microbial community models.

| Model | #metabolites* | #reactions* | $|\mathbf{J}^{tic}|$ | Total CPU time (sec)^ | | | Max. diff.& |
|---|---|---|---|---|---|---|---|
| | | | | Fast-SNP | NS-LLC | EFM-LLC | |
| 9-species model | 5758 | 7621 | 253 | 13512 ± 3683 | 1631 ± 39 | 1549 ± 26 | 8.7 x 10$^{-7}$ |
| Ec(2) | 2460 | 3615 | 130 | 998 ± 41 | 171 ± 1 | 161 ± 15 | 1 x 10$^{-8}$ |
| Ec(3) | 3625 | 5351 | 195 | 4538 ± 69 | 684 ± 20 | 621 ± 14 | 1.4 x 10$^{-5}$ |
| Ec(4) | 4790 | 7087 | 260 | 11807 ± 598 | 1404 ± 16 | 1267 ± 21 | 1.4 x 10$^{-5}$ |
| Ec(5) | 5955 | 8823 | 325 | 26927 ± 1708 | 2355 ± 60 | 1998 ± 64 | 1.9 x 10$^{-5}$ |
| Ec(6) | 7120 | 10559 | 390 | 48815 ± 3209 | 3614 ± 67 | 3011 ± 41 | 3.4 x 10$^{-5}$ |
| Ec(7) | 8285 | 12295 | 455 | 75971 ± 1940 | 5678 ± 271 | 4934 ± 37 | 3.5 x 10$^{-5}$ |
| Ec(8) | 9450 | 14031 | 520 | 125257 ± 7823 | 7695 ± 125 | 6531 ± 49 | 4.1 x 10$^{-5}$ |

Ec($n$) is the community model of $n$ copies of the *E. coli i*JO1366 model. $|\mathbf{J}^{tic}|$ is the number of reactions in TICs.
*Numbers of metabolites and reactions after removal of blocked reactions.
^ Mean ± s.d. of three replications.
&The maximum absolute difference of all tested methods compared to Fast-SNP for each model.

SI Methods

# Accelerating flux balance calculations in genome-scale metabolic models by localizing the application of loopless constraints

Siu H. J. Chan, Lin Wang, Satyakam Dash , Costas D. Maranas
Department of Chemical Engineering, The Pennsylvania State University

## Table of Contents

# 1 Notations

| | |
|---|---|
| **I** | Set of metabolites |
| **J** | Set of reactions |
| $\mathbf{J}^{int}$ | Set of internal reactions (mass-balanced reactions with substrate and products) |
| $\mathbf{J}^{ext}$ | Set of internal reactions (sink or source reactions that represent system input/output) |
| $\mathbf{S} = [S_{ij}]_{i \in \mathbf{I}, j \in \mathbf{J}}$ | Stoichiometric matrix |
| $\mathbf{S}^{int} = [S_{ij}]_{i \in \mathbf{I}, j \in \mathbf{J}^{int}}$ | Submatrix of **S** containing the columns for reactions in $\mathbf{J}^{int}$ |
| $\mathbf{S}^{ext} = [S_{ij}]_{i \in \mathbf{I}, j \in \mathbf{J}^{ext}}$ | Submatrix of **S** containing the columns for reactions in $\mathbf{J}^{ext}$ |
| $\mathbf{v} = [v_j]_{j \in \mathbf{J}}$ | Flux distribution satisfying the steady-state and bound constraints. |
| $\mathbf{v}^{int} = [v_j]_{j \in \mathbf{J}^{int}}$ | Vector containing the fluxes of all internal reactions |
| $\mathbf{v}^{ext} = [v_j]_{j \in \mathbf{J}^{ext}}$ | Vector containing the fluxes of all external reactions |
| $\mathbf{N}^{int} = [N_{jr}^{int}]_{j \in \mathbf{J}, r \in \{1,\ldots,R\}}$ | Nullspace matrix of $\mathbf{S}^{int}$ of rank $R$ |

# 2 Original loopless constraints

Loopless flux variability analysis (ll-FVA) with the original loopless constraints is restated as follows:

$$
\begin{aligned}
\min / \max \quad & v_j \\
\text{subject to} \quad & \sum_{j \in \mathbf{J}} S_{ij} v_j = 0, & \forall i \in \mathbf{I} & \quad \ldots\ldots\ldots(1) \\
& LB_j \leq v_j \leq UB_j, & \forall j \in \mathbf{J} & \quad \ldots\ldots\ldots(2) \\
& -M(1-y_j) \leq v_j \leq My_j, & \forall j \in \mathbf{J}^{int} & \quad \ldots\ldots\ldots(3) \\
& -My_j + \varepsilon(1-y_j) \leq G_j \leq -\varepsilon y_j + M(1-y_j), & \forall j \in \mathbf{J}^{int} & \quad \ldots\ldots\ldots(4) \\
& \sum_{j \in \mathbf{J}^{int}} N_{jr}^{int} G_j = 0, & \forall r \in \{1,\ldots,R\} & \quad \ldots\ldots\ldots(5) \\
& y_j \in \{0,1\}, & \forall j \in \mathbf{J}^{int} & \\
& G_j \in \mathbb{R}, & \forall j \in \mathbf{J}^{int} & \\
& v_j \in \mathbb{R}, & \forall j \in \mathbf{J} &
\end{aligned}
$$

where $LB_j$ and $UB_j$ are the lower and upper bounds for reaction $j$ respectively, $G_j$ is a continuous variable associated with internal reaction $j$ analogous to the change in Gibbs free energy of the reaction, $M$ is a large positive constant, $y_j$ is a binary variable to impose that $G_j$ must have an opposite sign to $v_j$ if $v_j$ is nonzero.

# 3    Definitions

The definitions are recited preceding the proofs for related propositions.

**Definition 1**

A TIC is a nonzero flux distribution **v** such that $\mathbf{S}^{int}\mathbf{v}^{int} = \mathbf{0}$.

**Definition 2**

A flux distribution **v** is said to contain another flux distribution $\mathbf{v}'$, denoted by $\mathbf{v}' \preceq \mathbf{v}$, if each flux $v'_j$ in $\mathbf{v}'$ is either zero or having the same sign as $v_j$ with a small or equal magnitude, i.e.,

$$0 \leq \operatorname{sgn}(v_j) v'_j \leq \operatorname{sgn}(v_j) v_j, \quad \forall j \in \mathbf{J}$$

where $\operatorname{sgn}(v_j) = \begin{cases} 1 & \text{if } v_j \geq 0 \\ -1 & \text{otherwise} \end{cases}$.

Furthermore, **v** is said to properly contain $\mathbf{v}'$, denoted by $\mathbf{v}' \prec \mathbf{v}$, if in addition to $\mathbf{v}' \preceq \mathbf{v}$ there is at least one reaction having nonzero flux in **v** but zero flux in $\mathbf{v}'$, i.e.,

$$\exists j' \in \mathbf{J} \text{ s.t. } v'_{j'} = 0 \text{ and } v_j \neq 0$$

**Definition 3**

A flux distribution **e** is an elementary flux mode (EFM) if $\mathbf{v} \prec \mathbf{e} \Rightarrow \mathbf{v} = \mathbf{0}$.

**Definition 4**

Two reactions $j_1, j_2$ are connected under the nullspace matrix $\mathbf{N}^{int}$, denoted by $j_1 \sim j_2$, if there is $r \in \{1,\ldots,R\}$ such that $N^{int}_{j_1 r} N^{int}_{j_2 r} \neq 0$ or if there is $j_3$ such that $j_1 \sim j_3$ and $j_3 \sim j_2$.

Definition 4 is only verbally stated in the main text. The relation $\sim$ is an equivalence relation and thus the reactions and the associated columns in $\mathbf{N}^{int}$ can be partitioned into separate connected components. Assume there are $L$ connected components respectively for reactions in TICs $\mathbf{J}^{tic}_l \subset \mathbf{J}^{tic}$ for $l = 1,\ldots,L$ and columns in the nullspace matrix $\mathbf{R}_l \subset \{1,\ldots,R\}$ for $l = 1,\ldots,L$. This prepares us the context for proving the claim in section 2.5 in the main text. Please see the proof for Proposition 3 and Corollary 1 in this document.

# 4  Proofs for propositions

In the main text, the no-cancellation rule of decomposing flux distributions using EFMs is used. It is formally stated here with a proof provided. This rule was established and proved previously (Schuster *et al.*, 2002).

**No-cancellation rule**

$$\forall \mathbf{v}, \exists\, K \text{ EFMs } \mathbf{e}_1,\ldots,\mathbf{e}_K \preceq \mathbf{v} \text{ and } \alpha_1,\ldots,\alpha_K > 0 \text{ s.t. } \mathbf{v} = \sum_{k=1}^{K} \alpha_k \mathbf{e}_k$$

Proof:

If there is not any flux distribution $\mathbf{v}' \neq \mathbf{0}$ such that $\mathbf{v}' \prec \mathbf{v}$, then $\mathbf{v} = \mathbf{e}_1$ is already an EFM. If there is a flux distribution $\mathbf{v}'$ such that $\mathbf{v}' \prec \mathbf{v}$, then consider $\mathbf{v}'' = \mathbf{v} - \rho_{\mathbf{v}/\mathbf{v}'}\mathbf{v}'$ where $\rho_{\mathbf{v}/\mathbf{v}'} = v_{j*}/v'_{j*} = \min\{v_j/v'_j \mid j \in \mathbf{J},\, v'_j \neq 0\}$. Obviously, $\mathbf{S}\mathbf{v}'' = \mathbf{0}$. Note that

$$\begin{array}{ccccc}
\operatorname{sgn}(v_j)v_j - \operatorname{sgn}(v_j)(v_j/v'_j)v'_j & \leq & \operatorname{sgn}(v_j)v_j - \rho_{\mathbf{v}/\mathbf{v}'}\operatorname{sgn}(v_j)v'_j & \leq & \operatorname{sgn}(v_j)v_j - 0 \\
0 & \leq & \operatorname{sgn}(v_j)v''_j & \leq & \operatorname{sgn}(v_j)v_j
\end{array}$$

Since $v''_{j*} = v_{j*} - (v_{j*}/v'_{j*})v'_{j*} = 0$, this implies $\mathbf{v}'' \prec \mathbf{v}$. Iteratively applying the same argument results at an EFM $\mathbf{e}_1 \prec \mathbf{v}$. Let $\mathbf{v}_0 = \mathbf{v}$. Iteratively apply the procedure to the flux distribution $\mathbf{v}_k = \mathbf{v}_{k-1} - \rho_{\mathbf{v}_{k-1}/\mathbf{e}_k}\mathbf{e}_k$ to obtain $\mathbf{e}_{k+1} \prec \mathbf{v}$ for $k = 1, 2, \ldots$. Since there is a finite number of reactions and the number of nonzeros in $\mathbf{v}_k$ is at least less than that in $\mathbf{v}$ by $k$, there exists $K > 0$ such that $\mathbf{v}_{K-1} = \mathbf{v}_{K-2} - \rho_{\mathbf{v}_{K-2}/\mathbf{e}_{K-1}}\mathbf{e}_{K-1} = \mathbf{e}_K$ is an EFM. Therefore,

$$\mathbf{v} = \sum_{k=1}^{K-1} \rho_{\mathbf{v}_{k-1}/\mathbf{e}_k}\mathbf{e}_k + \mathbf{e}_K. \qquad\blacksquare$$

The 'no-cancellation' rule refers to the property that $\mathbf{v}$ contains all $\mathbf{e}_k$ decomposing it, i.e., all fluxes $e_{jk}$ for reaction $j$ in $\mathbf{e}_k$ have the same sign as $v_j$. Therefore there does not exist any two EFMs in the flux decomposition into EFMs that have fluxes with the opposite sign that can cancel each other.

Let $\mathbf{J}^{\text{tic}} = \{j \in \mathbf{J} \mid \exists\, \text{TIC } \mathbf{v} \text{ s.t. } v_j \neq 0\}$ be the set of reactions participating in any TICs. Let $\mathbf{K} = \{1,\ldots,K\}$ be the index set for all EFMs, $\mathbf{K}^{\text{ll}} = \{k \in \mathbf{K} \mid \exists\, j \in \mathbf{J}^{\text{ext}} \text{ s.t. } e_{jk} \neq 0\}$ be the set of loopless EFMs (EFMs with nonzero exchange fluxes do not contain any TIC by the minimality of EFMs). $\mathbf{K}^{\text{tic}} = \mathbf{K} \setminus \mathbf{K}^{\text{ll}}$ is then the set of all TIC EFMs since all exchange fluxes of the EFMs in $\mathbf{K}^{\text{tic}}$ are 0. Assume that we required that a flux distribution does not contain any TICs involving a given target set of reactions $\mathbf{T} \subset \mathbf{J}^{\text{tic}}$. Denote the set of TIC EFMs involving any reactions in $\mathbf{T}$ by $\mathbf{K}^{\text{tic}}_{\mathbf{T}} = \{k \in \mathbf{K}^{\text{tic}} \mid \exists\, j \in \mathbf{T} \text{ s.t. } e_{jk} \neq 0\}$. Define also $\mathbf{C}^{\text{EFM}}_{\mathbf{T}} = \{j \in \mathbf{J}^{\text{tic}} \mid \exists\, k \in \mathbf{K}^{\text{tic}}_{\mathbf{T}} \text{ s.t. } e_{jk} \neq 0\}$ as the set of reactions connected to $\mathbf{T}$ by any TIC EFMs. By the no-cancellation rule, $\mathbf{v}$ can be expressed as follows:

$$\begin{aligned}
\mathbf{v} &= \mathbf{v}^{ll} + \mathbf{v}^{tic} \\
&= \mathbf{v}^{ll} + \mathbf{v}^{tic,\,nontarget} + \mathbf{v}^{tic,\,target} \\
&= \sum_{k \in \mathbf{K}^{ll}} \alpha_k \mathbf{e}_k + \sum_{k \in \mathbf{K}^{tic} \setminus \mathbf{K}^{tic}_{\mathbf{T}}} \alpha_k \mathbf{e}_k + \sum_{k \in \mathbf{K}^{tic}_{\mathbf{T}}} \alpha_k \mathbf{e}_k, \quad \alpha_k \geq 0,\ \forall k \in \mathbf{K}
\end{aligned} \quad (6)$$

**Proposition 1**

A flux distribution $\mathbf{v}$ does not contain any TICs involving reactions in $\mathbf{T}$ if the following constraints are satisfied:

$$-M(1-y_j) \leq v_j \leq M y_j, \qquad \forall j \in \mathbf{C}^{EFM}_{\mathbf{T}} \qquad \ldots\ldots\ldots(7)$$

$$-M y_j + \varepsilon(1-y_j) \leq G_j \leq -\varepsilon y_j + M(1-y_j), \quad \forall j \in \mathbf{C}^{EFM}_{\mathbf{T}} \qquad \ldots\ldots\ldots(8)$$

$$\sum_{j \in \mathbf{J}^{int}} N^{int}_{jr} G_j = 0, \qquad \forall r \in \{1,\ldots,R\} \qquad \ldots\ldots\ldots(9)$$

$$g_j \in \mathbb{R}, \qquad \forall j \in \mathbf{J}^{int}$$

$$y_j \in \{0,1\}, \qquad \forall j \in \mathbf{C}^{EFM}_{\mathbf{T}}$$

Proof:

For a flux distribution $\mathbf{v}$, consider the following LP problem to find a maximal TIC $\mathbf{v}'^*$ involving reactions in $\mathbf{C}^{EFM}_{\mathbf{T}}$:

$$\begin{aligned}
\mathbf{v}'^* = \underset{v'_j,\, j \in \mathbf{J}^{int}}{\arg\max} & \quad \sum_{j \in \mathbf{J}^{int}} \mathrm{sgn}(v_j) v'_j \\
\text{subject to} & \quad \sum_{r=1}^{R} N^{int}_{jr} w_r = v'_j, \qquad \forall j \in \mathbf{J}^{int} \\
& \quad v'_j = 0, \qquad \forall j \in \mathbf{J}^{int} \setminus \mathbf{C}^{EFM}_{\mathbf{T}} \\
& \quad 0 \leq \mathrm{sgn}(v_j) v'_j \leq \mathrm{sgn}(v_j) v_j \quad \forall j \in \mathbf{J}^{int}
\end{aligned}$$

$\mathbf{v}'^*$ is well defined since $\mathbf{0}$ is always feasible. Clearly $\mathbf{v}'^* \preceq \mathbf{v}$ and $\mathbf{v} - \mathbf{v}'^* \preceq \mathbf{v}$. If $\mathbf{v} - \mathbf{v}'^*$ contains any TIC $\mathbf{v}''$ involving $\mathbf{T}$, i.e., $v''_j \neq 0$ for some $j \in \mathbf{T}$, this implies that there exists $k \in \mathbf{K}^{tic}_{\mathbf{T}}$ and $\alpha_k > 0$ such that $\alpha_k \mathbf{e}_k \preceq \mathbf{v}''$ by eq. (6). By the definitions of $\mathbf{K}^{tic}_{\mathbf{T}}$ and $\mathbf{C}^{EFM}_{\mathbf{T}}$, $\alpha_k \mathbf{e}_k$ is feasible to the LP, violating that $\mathbf{v}'^*$ is maximal. This proves that $\mathbf{v} - \mathbf{v}'^*$ does not contain any TICs involving any reaction in $\mathbf{T}$. If $\mathbf{v}'^* = \mathbf{0}$, then $\mathbf{v}$ does not contain any TICs involving $\mathbf{J}^{tic}_{target}$. By LP duality and complementary slackness, the maximization problem can be transformed into:

$$\sum_{r=1}^{R} N_{jr}^{\text{int}} w_r = v'_j, \qquad \forall j \in \mathbf{J}^{\text{int}} \qquad \text{(dual variable: } g_j\text{)}$$

$$v'_j = 0, \qquad \forall j \in \mathbf{J}^{\text{int}} \setminus \mathbf{C}_{\mathbf{T}}^{\text{EFM}} \qquad \text{(dual variable: } \lambda_j\text{)}$$

$$-\text{sgn}(v_j)v'_j \leq 0, \qquad \forall j \in \mathbf{J}^{\text{int}} \qquad \text{(dual variable: } \mu_j^L\text{)}$$

$$\text{sgn}(v_j)v'_j \leq \text{sgn}(v_j)v_j, \qquad \forall j \in \mathbf{J}^{\text{int}} \qquad \text{(dual variable: } \mu_j^U\text{)}$$

$$\sum_{j \in \mathbf{J}^{\text{int}}} N_{jr}^{\text{int}} g_j = 0, \qquad \forall r = 1,\ldots,R \qquad \text{(dual constraint for } w_r\text{)}$$

$$-g_j + \sum_{j' \in \mathbf{J}^{\text{int}} \setminus \mathbf{J}_{\text{target}}^{\text{tic}}} \delta_j^{j'} \lambda_{j'} - \text{sgn}(v_j)\mu_j^L + \text{sgn}(v_j)\mu_j^U = \text{sgn}(v_j), \qquad \forall j \in \mathbf{J}^{\text{int}} \qquad \text{(dual constraint for } v'_j\text{)}$$

$$v'_j \mu_j^L = 0, \qquad \forall j \in \mathbf{J}^{\text{int}} \qquad \text{(complementary slackness)}$$

$$(v_j - v'_j)\mu_j^U = 0, \qquad \forall j \in \mathbf{J}^{\text{int}} \qquad \text{(complementary slackness)}$$

$$\mu_j^L, \mu_j^U \geq 0, \qquad \forall j \in \mathbf{J}^{\text{int}} \qquad \text{(dual feasibility)}$$

$$g_j \in \mathbb{R}, \qquad \forall j \in \mathbf{J}^{\text{int}}$$

$$\lambda_j \in \mathbb{R}, \qquad \forall j \in \mathbf{J}^{\text{int}} \setminus \mathbf{C}_{\mathbf{T}}^{\text{EFM}}$$

where $\delta_j^{j'} = \begin{cases} 1 & \text{if } j' = j \\ 0 & \text{otherwise} \end{cases}$. Substituting $\mathbf{v}' = \mathbf{v}'^* = \mathbf{0}$, we have:

$$\sum_{j \in \mathbf{J}^{\text{int}}} N_{jr}^{\text{int}} g_j = 0, \qquad \forall r = 1,\ldots,R$$

$$g_j = \text{sgn}(v_j)(\mu_j^U - \mu_j^L - 1), \qquad \forall j \in \mathbf{C}_{\mathbf{T}}^{\text{EFM}}$$

$$g_j = \text{sgn}(v_j)(\mu_j^U - \mu_j^L - 1) + \lambda_j, \qquad \forall j \in \mathbf{J}^{\text{int}} \setminus \mathbf{C}_{\mathbf{T}}^{\text{EFM}}$$

$$v_j \mu_j^U = 0, \qquad \forall j \in \mathbf{J}^{\text{int}}$$

$$\mu_j^L, \mu_j^U \geq 0, \qquad \forall j \in \mathbf{J}^{\text{int}}$$

$$g_j \in \mathbb{R}, \qquad \forall j \in \mathbf{J}^{\text{int}}$$

$$\lambda_j \in \mathbb{R}, \qquad \forall j \in \mathbf{J}^{\text{int}} \setminus \mathbf{C}_{\mathbf{T}}^{\text{EFM}}$$

Note that $g_j$ is free $\forall j \in \mathbf{J}^{\text{int}} \setminus \mathbf{C}_{\mathbf{T}}^{\text{EFM}}$ since $\lambda_j$ is unconstrained.

For $j \in \mathbf{C}_{\mathbf{T}}^{\text{EFM}}$:

- If $v_j > 0$, then $\mu_j^U = 0$ and $g_j = -\mu_j^L - 1 \leq -1$.
- If $v_j < 0$, then $\mu_j^U = 0$ and $g_j = \mu_j^L + 1 \geq 1$.
- If $v_j = 0$, then $g_j = \mu_j^U - (\mu_j^L + 1)$, which can take any real values.

The conditions can be transformed into exactly the same constraints stated in the proposition given a sufficiently large *M*. ∎

The general LP problem considered in the main text for finding a flux distribution **v**:

$$\min_{v_j, j \in \mathbf{J}} \sum_{j \in \mathbf{J}} c_j v_j$$

$$\text{subject to} \sum_{j \in \mathbf{J}} S_{ij} v_j = 0, \quad \forall i \in \mathbf{I}$$

$$\sum_{j \in \mathbf{J}} a_{pj} v_j \leq b_p, \quad \forall p \in \{1, \ldots, P\} \tag{10}$$

where $P$ is the number of additional constraints. Define two subsets of reactions in TICs: $\mathbf{J}_{fwd}^{tic} = \{j \in \mathbf{J}^{tic} \mid \exists k \in \mathbf{K}^{tic} \text{ s.t. } e_{jk} > 0\}$ and $\mathbf{J}_{rev}^{tic} = \{j \in \mathbf{J}^{tic} \mid \exists k \in \mathbf{K}^{tic} \text{ s.t. } e_{jk} < 0\}$ for reactions whose forward and reverse directions participating in TICs respectively. Note that $\mathbf{J}^{tic} = \mathbf{J}_{fwd}^{tic} \cup \mathbf{J}_{rev}^{tic}$. Proposition 3 establishes a useful sufficient condition for using only the localized loopless constraints:

**Proposition 2**
For the LP problem (10), assume that the target reaction set $\mathbf{T} \subset \mathbf{J}^{tic}$ contains all reactions in TICs satisfying one of the following three conditions:
(I)    $j \in \mathbf{J}_{fwd}^{tic}$ and $c_j < 0$
(II)   $j \in \mathbf{J}_{rev}^{tic}$ and $c_j > 0$
(III)  Both of the sub conditions (A) and (B) are satisfied:
   (A)    $(j \in \mathbf{J}_{fwd}^{tic}$ and $\exists p$ s.t. $a_{pj} < 0)$ or $(j \in \mathbf{J}_{rev}^{tic}$ and $\exists p$ s.t. $a_{pj} > 0)$
   (B)    $\exists p$ s.t. $a_{pj} \neq 0$ and $(a_{pj'} \neq 0$ for some $j' \neq j$ or $b_p < 0)$

Then the optimal objective function value of the LP problem constrained with the localized loopless constraints eqs. (7) – (9) is equal to the optimal objective function value of the LP problem constrained with the original loopless constraints eqs. (3) – (5).

Proof:
Let **v** be the optimal solution to the LP problem constrained with the localized loopless constraints. By eq. (6) and Proposition 2, **v** can be expressed as

$$\mathbf{v} = \mathbf{v}^{ll} + \mathbf{v}^{tic}$$
$$= \mathbf{v}^{ll} + \mathbf{v}^{tic, nontarget}$$
$$= \sum_{k \in \mathbf{K}^{ll}} \alpha_k \mathbf{e}_k + \sum_{k \in \mathbf{K}^{tic} \setminus \mathbf{K}_T^{tic}} \alpha_k \mathbf{e}_k, \quad \alpha_k \geq 0, \forall k \in \mathbf{K}$$

In what follows, we argue that the loopless part $\mathbf{v}^{ll}$ is feasible to the original LP with the same objective function value. Considering the $p$-th additional constraint:

$$\sum_{j \in \mathbf{J}} a_{pj} v_j \leq b_p$$

$$\sum_{j \in \mathbf{J}} a_{pj} v_j^{ll} + \sum_{j \in \mathbf{J}^{tic} \setminus \mathbf{J}_{target}^{tic}} a_{pj} v_j^{tic, nontarget} \leq b_p$$

Reactions in $\mathbf{J}^{tic} \setminus \mathbf{T}$ must violate either condition (III)(A) or (III)(B). Assume condition (III)(A) is violated for reaction $j \in \mathbf{J}^{tic} \setminus \mathbf{T}$.

- If $j \in \mathbf{J}_{\text{fwd}}^{\text{tic}} \setminus \mathbf{J}_{\text{rev}}^{\text{tic}}$, then $a_{pj} \geq 0 \ \forall p$.
- If $j \in \mathbf{J}_{\text{rev}}^{\text{tic}} \setminus \mathbf{J}_{\text{fwd}}^{\text{tic}}$, then $a_{pj} \leq 0 \ \forall p$.
- If $j \in \mathbf{J}_{\text{fwd}}^{\text{tic}} \cap \mathbf{J}_{\text{rev}}^{\text{tic}}$, then $a_{pj} = 0 \ \forall p$.

In any of the three cases, $a_{pj} v_j^{\text{tic, nontarget}} \geq 0 \ \forall p$.

Assume condition (III)(B) is violated for reaction $j \in \mathbf{J}^{\text{tic}} \setminus \mathbf{T}$. Then we have
$$a_{pj} = 0 \text{ or } (a_{pj'} = 0 \ \forall j' \neq j \text{ and } b_p \geq 0) \ \forall p$$

For each constraint $p$, if $a_{pj} = 0$, then $a_{pj} v_j^{\text{tic, nontarget}} = 0$. If $a_{pj'} = 0 \ \forall j' \neq j$ and $b_p \geq 0$, then the constraint becomes
$$a_{pj}(v_j^{\text{ll}} + v_j^{\text{tic, nontarget}}) \leq b_p$$

If $v_j^{\text{ll}} = 0$, then $a_{pj} v_j^{\text{ll}} \leq b_p$ since $b_p \geq 0$.

If $v_j^{\text{ll}} \neq 0$, then let $\kappa = v_j^{\text{tic, nontarget}} / v_j^{\text{ll}}$. $\kappa \geq 0$ as $\mathbf{v}^{\text{ll}} \preceq \mathbf{v}$ and $\mathbf{v}^{\text{tic, nontarget}} \preceq \mathbf{v}$, preserving the sign. Therefore,
$$a_{pj}(v_j^{\text{ll}} + \kappa v_j^{\text{ll}}) \leq b_p$$
$$a_{pj} v_j^{\text{ll}} \leq \frac{b_p}{1+\kappa} \leq b_p$$

Since all reactions in $\mathbf{J}^{\text{tic}} \setminus \mathbf{T}$ violate either condition (III)(A) or (III)(B), we have $\sum_{j \in \mathbf{J}} a_{pj} v_j^{\text{ll}} \leq b_p$ for all constraint $p$. $\mathbf{v}^{\text{ll}}$ is thus a feasible loopless solution to LP problem (10), which also satisfies the original loopless constraints.

For the optimality of $\mathbf{v}^{\text{ll}}$, since all reactions in $\mathbf{J}^{\text{tic}} \setminus \mathbf{T}$ violate conditions (I) and (II),

$j \in (\mathbf{J}^{\text{tic}} \setminus \mathbf{T}) \cap \mathbf{J}_{\text{fwd}}^{\text{tic}} \Rightarrow c_j \geq 0$ and $v_j^{\text{tic, nontarget}} \geq 0 \Rightarrow c_j v_j^{\text{tic, nontarget}} \geq 0$

$j \in (\mathbf{J}^{\text{tic}} \setminus \mathbf{T}) \cap \mathbf{J}_{\text{rev}}^{\text{tic}} \Rightarrow c_j \leq 0$ and $v_j^{\text{tic, nontarget}} \leq 0 \Rightarrow c_j v_j^{\text{tic, nontarget}} \geq 0$

$j \in \mathbf{J} \setminus (\mathbf{J}^{\text{tic}} \setminus \mathbf{T}) \Rightarrow v_j^{\text{tic, nontarget}} = 0 \Rightarrow c_j v_j^{\text{tic, nontarget}} = 0$

Therefore $\sum_{j \in \mathbf{J}} c_j v_j^{\text{tic, nontarget}} \geq 0$. By the optimality of $\mathbf{v}$, $\sum_{j \in \mathbf{J}} c_j v_j^{\text{tic, nontarget}} = 0$. Otherwise $\mathbf{v}^{\text{ll}}$ alone which is also feasible to LP problem (10) constrained with the localized loopless constraints eqs. (7) – (9) will have a lower objective function value than $\mathbf{v}$. Therefore $\sum_{j \in \mathbf{J}} c_j v_j = \sum_{j \in \mathbf{J}} c_j v_j^{\text{ll}}$. Since the localized loopless constraints are relaxation of the original loopless constraints, the minimum objective function value of LP problem (10) constrained with the original loopless constraints is bounded below by the minimum objective function value of LP problem (10) constrained with the localized loopless constraints. Hence, $\mathbf{v}^{\text{ll}}$ is an optimal solution to LP problem (10) constrained with the original loopless constraints having the same objective function value as $\mathbf{v}$. ∎

The following Proposition 3 is not explicitly stated in the main text. It is only verbally discussed in the main text. The relation $\sim$ is the relation of connection defined in Definition 4. Proposition 3 states that disconnect between two sets of reactions in the nullspace matrix implies that there is no connecting EFM between them.

**Proposition 3**

$$j_1 \not\sim j_2 \Rightarrow \nexists k \in \mathbf{K}^{tic} \text{ s.t. } e_{j_1 k} e_{j_2 k} \neq 0.$$

Proof:
Upon row and column permutation, $\mathbf{N}^{int}$ can be rearranged into

$$\mathbf{N}^{int} = \begin{bmatrix} \mathbf{N}_1^{int} & \mathbf{0} & \cdots & & \mathbf{0} \\ \mathbf{0} & \mathbf{N}_2^{int} & \mathbf{0} & \cdots & \mathbf{0} \\ \vdots & \mathbf{0} & \ddots & \ddots & \vdots \\ \vdots & \vdots & \ddots & \ddots & \mathbf{0} \\ \mathbf{0} & \mathbf{0} & \cdots & \mathbf{0} & \mathbf{N}_L^{int} \end{bmatrix}$$

where $\mathbf{N}_l^{int}$ is the submatrix from rows $\mathbf{J}_l^{tic}$ and columns $\mathbf{R}_l$ of $\mathbf{N}^{int}$ for $l = 1,\ldots,L$. Any TIC $\mathbf{v}$ can be expressed as

$$\mathbf{v} = \begin{bmatrix} \mathbf{N}_1^{int} & \mathbf{0} & \cdots & & \mathbf{0} \\ \mathbf{0} & \mathbf{N}_2^{int} & \mathbf{0} & \cdots & \mathbf{0} \\ \vdots & \mathbf{0} & \ddots & \ddots & \vdots \\ \vdots & \vdots & \ddots & \ddots & \mathbf{0} \\ \mathbf{0} & \mathbf{0} & \cdots & \mathbf{0} & \mathbf{N}_L^{int} \end{bmatrix} \begin{bmatrix} \mathbf{w}_1 \\ \mathbf{w}_2 \\ \vdots \\ \mathbf{w}_L \end{bmatrix} = \begin{bmatrix} \mathbf{N}_1^{int}\mathbf{w}_1 \\ \mathbf{N}_2^{int}\mathbf{w}_2 \\ \vdots \\ \mathbf{N}_L^{int}\mathbf{w}_L \end{bmatrix} = \begin{bmatrix} \mathbf{N}_1^{int}\mathbf{w}_1 \\ \mathbf{0} \\ \vdots \\ \mathbf{0} \end{bmatrix} + \begin{bmatrix} \mathbf{0} \\ \mathbf{N}_2^{int}\mathbf{w}_2 \\ \vdots \\ \mathbf{0} \end{bmatrix} + \cdots + \begin{bmatrix} \mathbf{0} \\ \mathbf{0} \\ \vdots \\ \mathbf{N}_L^{int}\mathbf{w}_L \end{bmatrix}$$

Note that

$$\mathbf{S}^{int}\begin{bmatrix} \mathbf{N}_1^{int}\mathbf{w}_1 \\ \mathbf{0} \\ \vdots \\ \mathbf{0} \end{bmatrix} = \mathbf{S}^{int}\begin{bmatrix} \mathbf{0} \\ \mathbf{N}_2^{int}\mathbf{w}_2 \\ \vdots \\ \mathbf{0} \end{bmatrix} = \cdots = \mathbf{S}^{int}\begin{bmatrix} \mathbf{0} \\ \mathbf{0} \\ \vdots \\ \mathbf{N}_L^{int}\mathbf{w}_L \end{bmatrix} = \mathbf{0}$$

Each vector satisfies the steady-state condition, i.e., eq. (1), as well as the bound constraints, i.e., eq. (2), as $\mathbf{v}$ also satisfies eq. (2). For any two reactions $j_1$, $j_2$, if $j_1 \not\sim j_2$, assume $j_1 \in \mathbf{J}_{l_1}^{tic}, j_2 \in \mathbf{J}_{l_2}^{tic}, l_1 \neq l_2$. If both reactions have nonzero fluxes in any flux distribution $\mathbf{v}$, then the two fluxes can always be separated into two flux distributions

$$\mathbf{v}_{l_1} = \begin{bmatrix} \mathbf{0} \\ \vdots \\ \mathbf{N}_{l_1}^{int}\mathbf{w}_{l_1} \\ \vdots \\ \mathbf{0} \end{bmatrix} \prec \mathbf{v}, \ \mathbf{v}_{l_2} = \begin{bmatrix} \mathbf{0} \\ \vdots \\ \mathbf{N}_{l_2}^{int}\mathbf{w}_{l_2} \\ \vdots \\ \mathbf{0} \end{bmatrix} \prec \mathbf{v}$$

Hence, there does not exist a single EFM with nonzero entries for both reactions. ∎

An immediate consequence of Proposition 3 is that a union of connected components that covers **T** also covers $\mathbf{C}_\mathbf{T}^{EFM}$. It is stated as Corollary 1.

**Corollary 1**

Let $\mathbf{C}_\mathbf{T}^{NS} = \bigcup_{q=1}^{Q} \mathbf{J}_{l_q}^{tic}$ where $1 \leq l_q \leq L$ for $q = 1,\ldots,Q$ such that $\mathbf{T} \subset \mathbf{C}_\mathbf{T}^{NS}$. Then $\mathbf{C}_\mathbf{T}^{EFM} \subset \mathbf{C}_\mathbf{T}^{NS}$.

Proof:

By Proposition 4, there does not exist any reaction in $\mathbf{J}^{tic} \setminus \mathbf{C}_\mathbf{T}^{NS}$ that shares an EFM with any reaction in **T**. Therefore, $\mathbf{C}_\mathbf{T}^{EFM} \subset \mathbf{C}_\mathbf{T}^{NS}$. ∎

# 5 Further relaxation of the localized loopless constraints

Regarding the localized loopless constraints stated in Proposition 2, if we have information on the directionality of reactions in $\mathbf{C}_\mathbf{T}^{\text{EFM}}$, it is possible to further discard some constraints in eqs. (7) – (9). However, in practical implementation, we found that the effect of the further relaxation is indefinite. Occasionally the computation appeared faster but more often it was slower, though the computational time was in the same order of magnitude. The further relaxation is presented as follows.

Define $\mathbf{C}_{\mathbf{T},f}^{\text{EFM}} = \{j \in \mathbf{J}^{\text{tic}} \mid \exists k \in \mathbf{K}_\mathbf{T}^{\text{tic}} \text{ s.t. } e_{jk} > 0\}$ and $\mathbf{C}_{\mathbf{T},r}^{\text{EFM}} = \{j \in \mathbf{J}^{\text{tic}} \mid \exists k \in \mathbf{K}_\mathbf{T}^{\text{tic}} \text{ s.t. } e_{jk} < 0\}$ for all reactions whose forward and reverse directions are respectively connected to $\mathbf{T}$ by any TIC EFM. Note that $\mathbf{C}_\mathbf{T}^{\text{EFM}} = \mathbf{C}_{\mathbf{T},f}^{\text{EFM}} \cup \mathbf{C}_{\mathbf{T},r}^{\text{EFM}}$. Then the localized loopless constraints can be further relaxed as follows:

$$\sum_{j \in \mathbf{J}^{\text{int}}} N_{jr}^{\text{int}} g_j = 0, \qquad \forall r \in \{1, \ldots, R\}$$

$$v_j \leq M z_j \qquad \forall j \in \mathbf{C}_{\mathbf{T},f}^{\text{EFM}} \quad \ldots\ldots\ldots\ldots(11)$$

$$g_j \leq -(M+1)z_j + M \quad \forall j \in \mathbf{C}_{\mathbf{T},f}^{\text{EFM}} \quad \ldots\ldots\ldots\ldots(12)$$

$$-M(1-z_j) \leq v_j \qquad \forall j \in \mathbf{C}_{\mathbf{T},r}^{\text{EFM}} \quad \ldots\ldots\ldots\ldots(13)$$

$$-(M+1)z_j + 1 \leq g_j \quad \forall j \in \mathbf{C}_{\mathbf{T},r}^{\text{EFM}} \quad \ldots\ldots\ldots\ldots(14)$$

$$z_j \in \{0,1\}, \qquad \forall j \in \mathbf{C}_\mathbf{T}^{\text{EFM}}$$

$$g_j \in \mathbb{R}, \qquad \forall j \in \mathbf{J}^{\text{int}}$$

Proof:

In the similar fashion, consider the following LP problem to find a maximal TIC $\mathbf{v}'^*$ involving reactions in $\mathbf{T}$:

$$\mathbf{v}'^* = \underset{v'_j, j \in \mathbf{J}^{\text{int}}}{\arg\max} \sum_{j \in \mathbf{J}^{\text{int}}} \text{sgn}(v_j) v'_j$$

$$\text{subject to } \sum_{r=1}^{R} N_{jr}^{\text{int}} w_r = v'_j, \qquad \forall j \in \mathbf{J}^{\text{int}}$$

$$v'_j = 0, \qquad \forall j \in \mathbf{J}^{\text{int}} \setminus \mathbf{C}_\mathbf{T}^{\text{EFM}}$$

$$v'_j \geq 0, \qquad \forall j \in \mathbf{C}_\mathbf{T}^{\text{EFM}} \setminus \mathbf{C}_{\mathbf{T},r}^{\text{EFM}}$$

$$v' \leq 0, \qquad \forall j \in \mathbf{C}_\mathbf{T}^{\text{EFM}} \setminus \mathbf{C}_{\mathbf{T},f}^{\text{EFM}}$$

$$0 \leq \text{sgn}(v_j) v'_j \leq \text{sgn}(v_j) v_j \quad \forall j \in \mathbf{J}^{\text{int}}$$

We can have the two additional constraints because all EFMs in $\mathbf{K}_\mathbf{T}^{\text{tic}}$ allow only non-negative and non-positive fluxes for reactions in $\mathbf{C}_{\mathbf{T},f}^{\text{EFM}}$ and $\mathbf{C}_{\mathbf{T},r}^{\text{EFM}}$ respectively. Similarly, we have the following constraints:

|  |  | primal/dual variables for the constraint |
|---|---|---|
| $\sum_{r=1}^{R} N_{jr}^{\text{int}} w_r = v'_j,$ | $\forall j \in \mathbf{J}^{\text{int}}$ | $g_j$ |
| $v'_j = 0,$ | $\forall j \in \mathbf{J}^{\text{int}} \setminus \mathbf{C}_{\mathbf{T}}^{\text{EFM}}$ | $\lambda_j$ |
| $v'_j \geq 0,$ | $\forall j \in \mathbf{C}_{\mathbf{T}}^{\text{EFM}} \setminus \mathbf{C}_{\mathbf{T},\text{r}}^{\text{EFM}}$ | $\lambda_j^f$ |
| $v'_j \leq 0,$ | $\forall j \in \mathbf{C}_{\mathbf{T}}^{\text{EFM}} \setminus \mathbf{C}_{\mathbf{T},\text{f}}^{\text{EFM}}$ | $\lambda_j^r$ |
| $-\text{sgn}(v_j) v'_j \leq 0,$ | $\forall j \in \mathbf{J}^{\text{int}}$ | $\mu_j^L$ |
| $\text{sgn}(v_j) v'_j \leq \text{sgn}(v_j) v_j,$ | $\forall j \in \mathbf{J}^{\text{int}}$ | $\mu_j^U$ |
| $\sum_{j \in \mathbf{J}^{\text{int}}} N_{jr}^{\text{int}} g_j = 0,$ | $\forall r \in \{1, \ldots, R\}$ | $w_r$ |

$$-g_j + \sum_{j' \in \mathbf{J}^{\text{int}} \setminus \mathbf{J}_{\text{target}}^{\text{tic}}} \delta_j^{j'} \lambda_{j'}$$
$$- \sum_{j' \in \mathbf{J}^{\text{int}} \setminus \mathbf{J}_{\text{target, r}}^{\text{tic}}} \delta_j^{j'} \lambda_{j'}^f + \sum_{j' \in \mathbf{J}^{\text{int}} \setminus \mathbf{J}_{\text{target, f}}^{\text{tic}}} \delta_j^{j'} \lambda_{j'}^r \quad \forall j \in \mathbf{J}^{\text{int}} \quad v'_j$$
$$- \text{sgn}(v_j) \mu_j^L + \text{sgn}(v_j) \mu_j^U = \text{sgn}(v_j),$$

|  |  |  |
|---|---|---|
| $v'_j \lambda_j^f = 0,$ | $\forall j \in \mathbf{C}_{\mathbf{T}}^{\text{EFM}} \setminus \mathbf{C}_{\mathbf{T},\text{r}}^{\text{EFM}}$ |  |
| $v'_j \lambda_j^r = 0,$ | $\forall j \in \mathbf{C}_{\mathbf{T}}^{\text{EFM}} \setminus \mathbf{C}_{\mathbf{T},\text{f}}^{\text{EFM}}$ |  |
| $v'_j \mu_j^L = 0,$ | $\forall j \in \mathbf{J}^{\text{int}}$ |  |
| $(v_j - v'_j) \mu_j^U = 0,$ | $\forall j \in \mathbf{J}^{\text{int}}$ |  |
| $\mu_j^L, \mu_j^U \geq 0,$ | $\forall j \in \mathbf{J}^{\text{int}}$ |  |
| $\lambda_j^f \geq 0,$ | $\forall j \in \mathbf{C}_{\mathbf{T}}^{\text{EFM}} \setminus \mathbf{C}_{\mathbf{T},\text{r}}^{\text{EFM}}$ |  |
| $\lambda_j^r \geq 0,$ | $\forall j \in \mathbf{C}_{\mathbf{T}}^{\text{EFM}} \setminus \mathbf{C}_{\mathbf{T},\text{f}}^{\text{EFM}}$ |  |
| $v'_j, g_j \in \mathbb{R},$ | $\forall j \in \mathbf{J}^{\text{int}}$ |  |
| $w_r \in \mathbb{R},$ | $\forall r \in \{1, \ldots, R\}$ |  |
| $\lambda_j \in \mathbb{R},$ | $\forall j \in \mathbf{J}^{\text{int}} \setminus \mathbf{C}_{\mathbf{T}}^{\text{EFM}}$ |  |

Substituting $\mathbf{v}' = \mathbf{v}'^* = \mathbf{0}$, we have:

$$\sum_{j \in \mathbf{J}^{\text{int}}} N_{jr}^{\text{int}} g_j = 0, \qquad \forall r \in \{1,\ldots,R\}$$

$$g_j = \text{sgn}(v_j)(\mu_j^U - \mu_j^L - 1) + \lambda_j, \quad \forall j \in \mathbf{J}^{\text{int}} \setminus \mathbf{C}_{\mathbf{T}}^{\text{EFM}}$$

$$g_j = \text{sgn}(v_j)(\mu_j^U - \mu_j^L - 1) - \lambda_j^f, \quad \forall j \in \mathbf{C}_{\mathbf{T}}^{\text{EFM}} \setminus \mathbf{C}_{\mathbf{T,r}}^{\text{EFM}}$$

$$g_j = \text{sgn}(v_j)(\mu_j^U - \mu_j^L - 1) + \lambda_j^r, \quad \forall j \in \mathbf{C}_{\mathbf{T}}^{\text{EFM}} \setminus \mathbf{C}_{\mathbf{T,f}}^{\text{EFM}}$$

$$g_j = \text{sgn}(v_j)(\mu_j^U - \mu_j^L - 1), \qquad \forall j \in \mathbf{C}_{\mathbf{T,f}}^{\text{EFM}} \cap \mathbf{C}_{\mathbf{T,r}}^{\text{EFM}}$$

$$v_j \mu_j^U = 0 \qquad \forall j \in \mathbf{J}^{\text{int}}$$

$$\mu_j^L, \mu_j^U \geq 0, \qquad \forall j \in \mathbf{J}^{\text{int}}$$

$$\lambda_j^f \geq 0, \qquad \forall j \in \mathbf{C}_{\mathbf{T}}^{\text{EFM}} \setminus \mathbf{C}_{\mathbf{T,r}}^{\text{EFM}}$$

$$\lambda_j^r \geq 0, \qquad \forall j \in \mathbf{C}_{\mathbf{T}}^{\text{EFM}} \setminus \mathbf{C}_{\mathbf{T,f}}^{\text{EFM}}$$

Note that $g_j$ is free $\forall j \in \mathbf{J}^{\text{int}} \setminus \mathbf{C}_{\mathbf{T}}^{\text{EFM}}$ since $\lambda_j$ is unconstrained.

For $j \in \mathbf{C}_{\mathbf{T}}^{\text{EFM}} \setminus \mathbf{C}_{\mathbf{T,r}}^{\text{EFM}}$:

- If $v_j > 0$, then $\mu_j^U = 0$ and $g_j = -\mu_j^L - \lambda_j^f - 1 \leq -1$.
- If $v_j < 0$, then $\mu_j^U = 0$ and $g_j = (\mu_j^L + 1) - \lambda_j^f$, which can take any real values.
- If $v_j = 0$, then $g_j = \mu_j^U - (\mu_j^L + \lambda_j^f + 1)$, which can take any real values.

For $j \in \mathbf{C}_{\mathbf{T}}^{\text{EFM}} \setminus \mathbf{C}_{\mathbf{T,f}}^{\text{EFM}}$:

- If $v_j > 0$, then $\mu_j^U = 0$ and $g_j = \lambda_j^r - (\mu_j^L + 1)$, which can take any real values.
- If $v_j < 0$, then $\mu_j^U = 0$ and $g_j = \mu_j^L + \lambda_j^r + 1 \geq 1$.
- If $v_j = 0$, then $g_j = (\mu_j^U + \lambda_j^r) - (\mu_j^L + 1)$, which can take any real values.

For $j \in \mathbf{C}_{\mathbf{T,f}}^{\text{EFM}} \cap \mathbf{C}_{\mathbf{T,r}}^{\text{EFM}}$:

- If $v_j > 0$, then $\mu_j^U = 0$ and $g_j = -\mu_j^L - 1 \leq -1$.
- If $v_j < 0$, then $\mu_j^U = 0$ and $g_j = \mu_j^L + 1 \geq 1$.
- If $v_j = 0$, then $g_j = \mu_j^U - (\mu_j^L + 1)$, which can take any real values.

The conditions can be transformed into exactly the same constraints stated above.

# 6 Nullspace algorithm

In the article, it is stated that the whole set of reactions participating in TICs can be determined efficiently by solving the following MILP problem:

$$\min \sum_{j \in \mathbf{J}^{int}} \left(z_j^+ + z_j^-\right)$$

$$\text{subject to} \quad \sum_{j \in \mathbf{J}^{int}} S_{ij} v_j = 0, \qquad \forall i \in \mathbf{I}$$

$$-M\delta_j^L \leq v_j \leq M\delta_j^U, \qquad \forall j \in \mathbf{J}^{int}$$

$$v_j \geq \varepsilon - M(z_j^+ + 1 - \delta_j^U), \quad \forall j \in \mathbf{J}^{int}$$

$$v_j \leq -\varepsilon + M(z_j^- + 1 - \delta_j^L), \quad \forall j \in \mathbf{J}^{int}$$

$$v_j \in \mathbb{R}, \qquad \forall j \in \mathbf{J}^{int}$$

$$z_j^+ \geq 0, \qquad \forall j \in \mathbf{J}^{int}$$

$$z_j^- \geq 0, \qquad \forall j \in \mathbf{J}^{int}$$

$$z_j^+, z_j^- \in \{0,1\}, \qquad \forall j \in \mathbf{J}^{int} \text{ if } \delta_j^L = \delta_j^U = 1 \quad \text{..............(11)}$$

where $\delta_j^L = \begin{cases} 1 & \text{if } LB_j < 0 \\ 0 & \text{if } LB_j \geq 0 \end{cases}$ and $\delta_j^U = \begin{cases} 1 & \text{if } UB_j > 0 \\ 0 & \text{if } UB_j \leq 0 \end{cases}$ are parameters depending on reaction directionality, $\varepsilon$ is a small positive number and $M$ is a large positive number. It can formally stated as the following proposition.

**Proposition 4**

$j \in \mathbf{J}^{tic}$ if and only if reaction $j$ has a nonzero flux in an optimal solution to problem (11).

Proof:
First, rewrite problem (11) as follows:

$$\min \sum_{j \in \mathbf{J}^{int}} \left(z_j^+ + z_j^-\right)$$

$$\text{subject to} \quad \sum_{j \in \mathbf{J}^{int}} S_{ij} v_j = 0, \qquad \forall i \in \mathbf{I}$$

$$-M\delta_j^L \leq v_j \leq M\delta_j^U, \quad \forall j \in \mathbf{J}^{int}$$

$$v_j \geq \varepsilon - M z_j^+, \qquad \forall j \in \mathbf{J}^{int}, \text{ s.t. } UB_j > 0$$

$$v_j \leq -\varepsilon + M z_j^-, \qquad \forall j \in \mathbf{J}^{int}, \text{ s.t. } LB_j < 0$$

$$v_j \in \mathbb{R}, \qquad \forall j \in \mathbf{J}^{int}$$

$$z_j^+ \geq 0, \qquad \forall j \in \mathbf{J}^{int}, \text{ s.t. } LB_j \geq 0$$

$$z_j^- \geq 0, \qquad \forall j \in \mathbf{J}^{int}, \text{ s.t. } UB_j \leq 0$$

$$z_j^+, z_j^- \in \{0,1\}, \qquad \forall j \in \mathbf{J}^{int}, \text{ s.t. } UB_j > 0 \text{ and } LB_j < 0$$

Let $(\mathbf{v}^*, \mathbf{z}^{+,*}, \mathbf{z}^{-,*})$ be an optimal solution where $\mathbf{v}^*$ is the optimal flux distribution and $\mathbf{z}^{+,*}$, $\mathbf{z}^{-,*}$ are the vectors containing $z_j^{+,*}$, $z_j^{-,*}$ for each $j$. $\mathbf{v}^*$ would have been scaled to have either $|v_j^*| = 0$ or $|v_j^*| \geq \varepsilon$ such that the corresponding $\mathbf{z}^{+,*}, \mathbf{z}^{-,*}$ are minimized:

$$z_j^{+,*} = \begin{cases} 0 & \text{if } v_j^* \geq \varepsilon \\ \varepsilon/M & \text{if } v_j^* = 0 \end{cases} \quad \forall j \in \mathbf{J}, \; LB_j \geq 0$$

$$z_j^{-,*} = \begin{cases} 0 & \text{if } v_j^* \leq -\varepsilon \\ \varepsilon/M & \text{if } v_j^* = 0 \end{cases} \quad \forall j \in \mathbf{J}, \; UB_j \leq 0$$

$$z_j^{+,*} = \begin{cases} 0 & \text{if } v_j^* \geq \varepsilon \\ 1 & \text{if } v_j^* = 0 \end{cases} \quad \forall j \in \mathbf{J}, \; UB_j > 0 \text{ and } LB_j < 0$$

$$z_j^{-,*} = \begin{cases} 0 & \text{if } v_j^* \leq -\varepsilon \\ 1 & \text{if } v_j^* = 0 \end{cases} \quad \forall j \in \mathbf{J}, \; UB_j > 0 \text{ and } LB_j < 0$$

Now for any reaction $j_0$, assume there exists a TIC $\mathbf{v}$ such that $v_{j_0} > 0$ but $v_{j_0}^* = 0$, $z_{j_0}^{+,*} > 0$ and $z_{j_0}^{-,*} = 1$ if $z_{j_0}^{-,*}$ is defined (which means reaction $j_0$ is truly reversible with $UB_{j_0} > 0$ and $LB_{j_0} < 0$). Consider following flux distribution

$$\mathbf{v}' = \beta \mathbf{v}^* + \frac{\varepsilon}{v_{j_0}} \mathbf{v}$$

where $\beta = \max\left\{ \max\left\{ \left|\frac{v_{j_0} + v_j}{v_{j_0}}\right| \middle| j \in \mathbf{J}, v_j^* < 0, v_j > 0 \right\}, \max\left\{ \left|\frac{v_{j_0} - v_j}{v_{j_0}}\right| \middle| j \in \mathbf{J}, v_j^* > 0, v_j < 0 \right\}, 1 \right\}$.

For any reaction $j$ with $z_j^{+,*} = 0$ and therefore $v_j^* \geq \varepsilon$, if $v_j \geq 0$, obviously $v_j' \geq \varepsilon$ is still satisfied. Similarly, for any reaction $j$ with $z_j^{-,*} = 0$ and therefore $v_j^* \leq -\varepsilon$, if $v_j \leq 0$, $v_j' \leq -\varepsilon$ is still satisfied.

For reaction $j$ with $z_j^{+,*} = 0$, $v_j^* \geq \varepsilon$, if $v_j < 0$, we have

$$v_j' = \beta v_j^* + \frac{\varepsilon}{v_{j_0}} v_j \geq \left(\frac{v_{j_0} - v_j}{v_{j_0}}\right) \varepsilon + \frac{\varepsilon v_j}{v_{j_0}} = \varepsilon$$

For reaction $j$ with $z_j^{-,*} = 0$, $v_j^* \leq -\varepsilon$, if $v_j > 0$, we have

$$v_j' = \beta v_j^* + \frac{\varepsilon}{v_{j_0}} v_j \leq \left(\frac{v_{j_0} + v_j}{v_{j_0}}\right)(-\varepsilon) + \frac{\varepsilon v_j}{v_{j_0}} = -\varepsilon$$

Therefore $(\mathbf{v}', \mathbf{z}^{+,*}, \mathbf{z}^{-,*})$ is also a feasible solution to problem (11) provided a sufficiently large $M$. However, since $v'_{j_0} = \beta(0) + \frac{\varepsilon v_{j_0}}{v_{j_0}} = \varepsilon$, we can define $\mathbf{z}'^+$ such that

$$z'^+_j = \begin{cases} 0 & \text{if } j = j_0 \\ z^{+,*}_j & \text{otherwise} \end{cases}$$

Then $(\mathbf{v}', \mathbf{z}'^+, \mathbf{z}^{-,*})$ is another feasible solution with a smaller objective function value as $z^{+,*}_{j_0} > 0$, contradicting the optimality of $(\mathbf{v}^*, \mathbf{z}^{+,*}, \mathbf{z}^{-,*})$. Therefore an optimal flux distribution must have $v^*_{j_0} \neq 0$ if there exists a TIC $\mathbf{v}$ such that $v_{j_0} > 0$.

The analogous argument also applies to any reaction $j_0$ for which there exists a TIC $\mathbf{v}$ such that $v_{j_0} < 0$. Therefore an optimal flux distribution must have $v^*_{j_0} \neq 0$ if there exists a TIC $\mathbf{v}$ such that $v_{j_0} < 0$. ∎

One may wonder whether it is possible to also relax the binary variables $z^+_j, z^-_j$ to non-negative continuous variables for each truly reversible reaction $j$ and meanwhile get the same desired results. In general relaxing problem (11) to a completely linear problem does not guarantee that all reactions in TICs have nonzero fluxes. For example, in the argument in the proof, $z^{-,*}_{j_0}$ is no longer fixed at 1 but depends on the value of $v^*_{j_0}$ and a smaller objective function value cannot be guaranteed. Indeed,

$$z^+_{j_0} + z^-_{j_0} = \begin{cases} 0 + \dfrac{\varepsilon - v_j}{M} & \text{if } v_j \geq \varepsilon \\[6pt] \dfrac{\varepsilon + v_j}{M} + \dfrac{\varepsilon - v_j}{M} & \text{if } -\varepsilon \leq v_j \leq \varepsilon \\[6pt] \dfrac{\varepsilon + v_j}{M} + 0 & \text{if } v_j \leq \varepsilon \end{cases}$$

$$= \begin{cases} \dfrac{\varepsilon + v_j}{M} & \text{if } v_j \geq \varepsilon \\[6pt] \dfrac{2\varepsilon}{M} & \text{if } -\varepsilon \leq v_j \leq \varepsilon \\[6pt] \dfrac{\varepsilon - v_j}{M} & \text{if } v_j \leq -\varepsilon \end{cases}$$

Therefore, $v_j = 0$ for all reversible reaction $j$ would be one of the optimal solutions to the relaxed problem (11).

**Properties of the null-space algorithm leading to its fast solution**

In the main text, the properties key to the fast solution of problem (11) are briefly discussed. Here we discuss them in more details.

Problem (11) forces each reaction in TICs to have nonzero flux by minimizing $z_j^+ + z_j^-$. For reaction with $LB_j \geq 0$ (i.e., active only in the forward direction), $\delta_j^L = 0$ and $z_j^- = 0$ is always feasible because the only constraint involving $z_j^-$ becomes slack. Therefore, $z_j^- = 0$ will be taken for optimality. Similarly, for reaction with $UB_j \leq 0$ (i.e., active only in the reverse direction), $z_j^+ = 0$ will be taken for optimality. Thus, only one of the $z_j^+, z_j^-$ needs to be determined for each irreversible reaction $j$. Modeling $z_j^+, z_j^-$ as continuous variables suffices to force $v_j \neq 0$ if feasible.

For reversible reaction $j$, $z_j^+, z_j^-$ are required to be binary. If $z_j^+ = 0$, then $v_j > 0$. If $z_j^- = 0$, then $v_j < 0$. If $z_j^+ = z_j^- = 1$, which is always feasible, then $v_j$ is unconstrained. A special property of problem (11) leading to its fast solution is that when branching any $z_j^+$ (or $z_j^-$), if $z_j^+ = 0$ is feasible, then there is an optimal solution with $z_j^+ = 0$ and $z_j^- = 1$ since the corresponding TIC with $v_j > 0$ can always be superimposed with any TICs with $v_j = 0$ (from the proof for Proposition 4) and become a feasible solution with a smaller objective function value. Therefore, a depth-first search with branching down whenever possible (take $z_j^+$ or $z_j^- = 0$) for all reversible reaction $j$ can always lead to an optimal solution, i.e., at most $2n_{rev}$ relaxed LPs need to be solved for a network with $n_{rev}$ reversible reactions. Solving problem (11) is similar to an iterative LP procedure but the former takes advantage of the built-in search structure and information flow of any modern MILP solver.

# 7     **Practical implementation of the preprocessing steps**

In practice, to minimize computational cost, the nullspace algorithm and the determination of $\mathbf{J}_{fwd}^{tic}$ and $\mathbf{J}_{rev}^{tic}$ (Step 1 and Step 2 in the LLC procedure in section 2.6 in the main text) are executed in an interlaced fashion as follows:

1. Initialize $\mathbf{J}_{fwd}^{tic} = \mathbf{J}_{rev}^{tic} = \{\}$.
2. From problem (11), fix $z_j^+ = z_j^- = 1$ for all truly reversible reaction $j$ with $UB_j > 0$ and $LB_j < 0$. Solve the resultant LP problem to obtain an optimal flux distribution $\mathbf{v}$.
3. Update $\mathbf{J}_{fwd}^{tic}$ and $\mathbf{J}_{rev}^{tic}$:
   - $\mathbf{J}_{fwd}^{tic} \leftarrow \mathbf{J}_{fwd}^{tic} \cup \{j \in \mathbb{R} \mid v_j > 0\}$
   - $\mathbf{J}_{rev}^{tic} \leftarrow \mathbf{J}_{rev}^{tic} \cup \{j \in \mathbb{R} \mid v_j < 0\}$
4. Define the set $\mathbf{J}_{fwd, fix}^{tic} = \mathbf{J}_{fwd}^{tic} \cup \{j \in \mathbf{J}^{int} \mid LB_j \geq 0\}$. From problem (11), fix $z_j^+ = 1$, $\forall j \in \mathbf{J}_{fwd, fix}^{tic}$ and $z_j^- = 1$, $\forall j \in \mathbf{J}^{int}$. Relax $z_j^+$, $\forall j \notin \mathbf{J}_{fwd, fix}^{tic}$ to continuous variables and solve the resultant LP problem to obtain an optimal flux distribution $\mathbf{v}$. Update $\mathbf{J}_{fwd}^{tic}$ and $\mathbf{J}_{rev}^{tic}$ by repeating Step 3.
5. Define the set $\mathbf{J}_{rev, fix}^{tic} = \mathbf{J}_{rev}^{tic} \cup \{j \in \mathbf{J}^{int} \mid UB_j \leq 0\}$. From problem (11), fix $z_j^+ = 1$, $\forall j \in \mathbf{J}^{int}$ and $z_j^- = 1$, $\forall j \in \mathbf{J}_{rev, fix}^{tic}$. Relax $z_j^-$, $\forall j \notin \mathbf{J}_{rev, fix}^{tic}$ to continuous variables and solve the resultant LP problem to obtain an optimal flux distribution $\mathbf{v}$. Update $\mathbf{J}_{fwd}^{tic}$ and $\mathbf{J}_{rev}^{tic}$ by repeating Step 3.
6. Update the set $\mathbf{J}_{fwd, fix}^{tic} \leftarrow \mathbf{J}_{fwd}^{tic} \cup \{j \in \mathbf{J}^{int} \mid LB_j \geq 0\}$. From problem (11), fix $z_j^+ = 1$, $\forall j \in \mathbf{J}_{fwd, fix}^{tic}$ and $z_j^- = 1$, $\forall j \in \mathbf{J}^{int}$. Solve the resultant MILP problem to obtain an optimal flux distribution $\mathbf{v}$. Update $\mathbf{J}_{fwd}^{tic}$ and $\mathbf{J}_{rev}^{tic}$ by repeating Step 3.
7. For each reaction $j' \in \mathbf{J}_{fwd, check}^{tic} = \{j \mid v_j < 0\} \setminus \mathbf{J}_{fwd, fix}^{tic}$, check whether it can carry a positive flux by:
   a. If $j' \in \mathbf{J}_{fwd}^{tic}$, go to Step (b). Else from problem (11), fix $z_j^+ = 1$, $\forall j \neq j'$ and $z_{j'}^+ = 0$ and $z_j^- = 1$, $\forall j \in \mathbf{J}^{int}$. Solve the resultant LP problem. If it has a feasible flux distribution is $\mathbf{v}$, update $\mathbf{J}_{fwd}^{tic}$ and $\mathbf{J}_{rev}^{tic}$ by repeating Step 3.
   b. Remove $j'$ from $\mathbf{J}_{fwd, fix}^{tic}$.
   c. Repeat Step (a) and (b) until $\mathbf{J}_{fwd, fix}^{tic}$ is empty.
8. Update the set $\mathbf{J}_{rev, fix}^{tic} \leftarrow \mathbf{J}_{rev}^{tic} \cup \{j \in \mathbf{J}^{int} \mid UB_j \leq 0\}$. From problem (11), fix $z_j^+ = 1$, $\forall j \in \mathbf{J}^{int}$ and $z_j^- = 1$, $\forall j \in \mathbf{J}_{rev}^{tic} \cup \{j \in \mathbf{J}^{int} \mid UB_j \leq 0\}$. Solve the resultant MILP problem to obtain an optimal flux distribution $\mathbf{v}$. Update $\mathbf{J}_{fwd}^{tic}$ and $\mathbf{J}_{rev}^{tic}$ by repeating Step 3.

9. For each reaction $j' \in \mathbf{J}_{\text{rev, check}}^{\text{tic}} = \{j\,|\,v_j > 0\} \setminus \mathbf{J}_{\text{rev, fix}}^{\text{tic}}$, check whether it can carry a negative flux by:
   a. If $j' \in \mathbf{J}_{\text{rev}}^{\text{tic}}$, go to Step (b). Else from problem (11), fix $z_j^+ = 1$, $\forall j \in \mathbf{J}^{\text{int}}$ and $z_j^- = 1$, $\forall j \neq j'$ and $z_{j'}^- = 0$. Solve the resultant LP problem. If it has a feasible flux distribution is $\mathbf{v}$, update $\mathbf{J}_{\text{fwd}}^{\text{tic}}$ and $\mathbf{J}_{\text{rev}}^{\text{tic}}$ by repeating Step 3.
   b. Remove $j'$ from $\mathbf{J}_{\text{fwd, fix}}^{\text{tic}}$.
   c. Repeat Step (a) and (b) until $\mathbf{J}_{\text{fwd, fix}}^{\text{tic}}$ is empty.
10. The sets $\mathbf{J}_{\text{fwd}}^{\text{tic}}$ and $\mathbf{J}_{\text{rev}}^{\text{tic}}$ have been completed. Let $\mathbf{J}^{\text{tic}} = \mathbf{J}_{\text{fwd}}^{\text{tic}} \cup \mathbf{J}_{\text{rev}}^{\text{tic}}$. Compute a nullspace matrix $\mathbf{N}^{\text{int}}$ from the columns in the stoichiometric matrix $\mathbf{S}$ corresponding to the reactions in $\mathbf{J}^{\text{tic}}$, i.e., $[S_{ij}]_{i \in \mathbf{I}, j \in \mathbf{J}^{\text{tic}}}$.

Step 2 – 3 would already identify all irreversible reactions that are in TICs and some reversible reactions that are in TICs. Only reversible reactions that are not yet identified need to be checked whether they are in TICs. Step 4 – 5 solve two relaxed LP problems as heuristics to find TICs that have the sum of positive fluxes larger than the sum of negative fluxes and vice versa. This can reduce the number of binary variables in Step 6 and Step 8. After solving Step 6, the forward feasibility for reactions having negative fluxes in the solution to Step 6 is checked to ensure that $\mathbf{J}_{\text{fwd}}^{\text{tic}}$ is completed. (It is possible that when solving Step 6, a reversible TIC consisting of two reactions shows fluxes of $[1\ -1]^T$ but indeed $[-1\ 1]^T$ is also feasible.) Step 8 – 9 similarly complete $\mathbf{J}_{\text{rev}}^{\text{tic}}$.

## 8 Algorithm for finding the connected components of null-space

The following algorithm is used in Step 3 of the LLC procedure presented in section 2.6 in the main text to identify all connected components of reactions $\mathbf{J}_l^{\text{tic}} \subset \mathbf{J}^{\text{tic}}$ of the null-space. $R$ is the rank of the null-space matrix.

A. $l = 0$. $\mathbf{R}_{\text{unchecked}} = \{1, 2, \ldots, R\}$. Go to Step B.

B. If $\mathbf{R}_{\text{unchecked}}$ is empty, terminate the algorithm.
   Else $l \leftarrow l+1$. $\mathbf{J}_l^{\text{tic}} = \mathbf{R}_{\text{current}} = \{\}$.
   Move the first element in $\mathbf{R}_{\text{unchecked}}$ to $\mathbf{R}_{\text{current}}$. Go to Step C.

C. $\mathbf{J}_{l, \text{new}}^{\text{tic}} = \{j \notin \mathbf{J}_l^{\text{tic}}\,|\,\exists r \in \mathbf{R}_{\text{current}}\ \text{s.t.}\ N_{jr}^{\text{int}} \neq 0\}$.
   $\mathbf{J}_l^{\text{tic}} \leftarrow \mathbf{J}_l^{\text{tic}} \cup \mathbf{J}_{l, \text{new}}^{\text{tic}}$.
   $\mathbf{R}_{\text{new}} = \{r \in \mathbf{R}_{\text{unchecked}}\,|\,\exists j \in \mathbf{J}_{l, \text{new}}^{\text{tic}}\ \text{s.t.}\ N_{jr}^{\text{int}} \neq 0\}$.
   If $\mathbf{J}_{l, \text{new}}^{\text{tic}}$ or $\mathbf{R}_{\text{new}}$ is empty, go to Step B.
   Else $\mathbf{R}_{\text{current}} \leftarrow \mathbf{R}_{\text{new}}$. $\mathbf{R}_{\text{unchecked}} \leftarrow \mathbf{R}_{\text{unchecked}} \setminus \mathbf{R}_{\text{new}}$. Repeat Step C.